\documentclass[10pt]{article}

\usepackage{amsmath}
\usepackage{amssymb}
\usepackage{amsfonts}
\usepackage{graphics}
\usepackage{MnSymbol}

\usepackage{duotenzor}

\title{\textbf{A formalism-local framework for general probabilistic theories including quantum theory}}
\author{Lucien Hardy\\
\textit{Perimeter Institute,}\\
\textit{31 Caroline Street North,}\\
\textit{Waterloo, Ontario N2L 2Y5, Canada}}

\begin{document}

\maketitle


\hspace{-8mm}\makebox[\textwidth]{\bf Abstract}
\vspace{8pt}

{\small{
\noindent\hspace{7mm}\makebox[10.7cm][s]{In this paper we consider general probabilistic theories that pertain to cir-}
\par
\noindent\hspace{7mm}\makebox[10.7cm][s]{cuits which satisfy two very natural assumptions.  We provide a formalism}
\par
\noindent\hspace{7mm}\makebox[10.7cm][s]{that is local in the following very specific sense: calculations pertaining to}
\par
\noindent\hspace{7mm}\makebox[10.7cm][s]{any region of spacetime employ only mathematical objects associated with}
\par
\noindent\hspace{7mm}\makebox[10.7cm][s]{that region. We call this {\it formalism locality}.  It incorporates the idea that}
\par
\noindent\hspace{7mm}\makebox[10.7cm][s]{space and time should be treated on an equal footing.  Formulations that use}
\par
\noindent\hspace{7mm}\makebox[1.2cm][l]
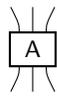
{
\begin{Diagram}{0}{0}
\boundingbox{0,0}{0,0}
\begin{move}[0.9]{2,-3.1}
\Opbox{A}{0,0}
\inwire[-5]{A}{1}\inwire{A}{2}\inwire[5]{A}{3}
\outwire[-5]{A}{1}\outwire{A}{2}\outwire[5]{A}{3}
\end{move}
\end{Diagram} }\hspace{-1.05mm}
\makebox[7.9cm][s]{a foliation of spacetime to evolve a state do not have this}
\makebox[1.7cm][l]{
\begin{Diagram}[0.7]{0}{0}
\boundingbox{0,0}{0,0}
\begin{move}{2.2,-11.2}
\Opbox{A}{0,0} \Opbox[2]{B}{3,4} \Opbox[2]{C}{1,8}
\wire{A}{B}{3}{1}\wire{B}{C}{1}{2}
\inwire[-5]{A}{1}\inwire{A}{2}\inwire[5]{A}{3}
\outwire[-5]{A}{1}\outwire{A}{2}
\inwire[5]{B}{2}\outwire[5]{B}{2}
\inwire[-5]{C}{1} \outwire{C}{1.5}
\end{move}
\end{Diagram} }
\par
\noindent\hspace{7mm}\makebox[1.2cm][r]{}\makebox[7.9cm][s]{property nor do histories-based approaches. An opera-}
\par
\noindent\hspace{7mm}\makebox[1.2cm][r]{}\makebox[7.9cm][s]{tion (see figure on left) has inputs and outputs (through}
\par
\noindent\hspace{7mm}\makebox[1.2cm][r]{}\makebox[7.9cm][s]{which systems travel). A circuit is built by wiring}
\par
\noindent\hspace{7mm}\noindent\makebox[9cm][s]{together operations such that we have no open inputs or outputs}
\par
\noindent\hspace{7mm}\noindent\makebox[9.1cm][s]{left over.  A fragment (see figure on right) is a part of a circuit}
\par
\noindent\hspace{7mm}\noindent\makebox[9.1cm][s]{and may have open inputs and outputs. We show how each}
\par
\noindent\hspace{7mm}\noindent\makebox[10.7cm][s]{operation is associated with a certain mathematical object which we call}
\par
\noindent\hspace{7mm}\noindent\makebox[10.7cm][s]{a {\it duotensor} (this is like a tensor but with a  bit more structure). In the}
\par
\noindent\hspace{7mm}\makebox[1.5cm][s]
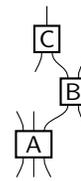
{
\begin{Diagram}{0}{0}
\boundingbox{0,0}{0,0}
\begin{move}[0.85]{-0.1,-2.3}
\Duobox{A}{0,0}
\inblack[-5]{A}{1}\inwhite{A}{2}\inwhite[5]{A}{3}
\outblack[-5]{A}{1}\outwhite{A}{2}\outblack[5]{A}{3}
\end{move}
\end{Diagram}
}\hspace{-1.05mm}
\makebox[9.2cm][s]{figure on the left we show how a duotensor is represented}
\par
\noindent\hspace{7mm}\makebox[1.5cm][r]{}\makebox[9.2cm][s]{graphically. We can link duotensors together such that black and}
\par
\noindent\hspace{7mm}\makebox[1.5cm][r]{}\makebox[6.4cm][s]{white dots match up to get the duotensor}
\par
\noindent\hspace{7mm}\makebox[7.9cm][s]{corresponding to any fragment.  The figure on the right}
\makebox[2.8cm][l]{
\begin{Diagram}[0.75]{0}{0}
\boundingbox{0,0}{0,0}
\begin{move}{2.6,-0.6}
\Duobox{A}{0,0} \Duobox[2]{B}{4,-3} \Duobox[2]{C}{8,-1}
\link{A}{B}{3}{1}\link{B}{C}{1}{2}
\inblack[-5]{A}{1}\inwhite{A}{2}\inwhite[5]{A}{3}
\outblack[-5]{A}{1}\outwhite{A}{2}
\inblack[5]{B}{2}\outwhite[5]{B}{2}
\inwhite[-5]{C}{1} \outblack{C}{1.5}
\end{move}
\end{Diagram}
}
\par
\noindent\hspace{7mm}\makebox[7.9cm][s]{is the duotensor for the above fragment.  Links represent}
\par
\noindent\hspace{7mm}\makebox[7.9cm][s]{summing over the corresponding indices.  We can use}
\par
\noindent\hspace{7mm}\makebox[10.7cm][s]{such duotensors to make probabilistic statements pertaining to fragments.}
\par
\noindent\hspace{7mm}\makebox[10.7cm][s]{Since fragments are the circuit equivalent of arbitrary spacetime regions}
\par
\noindent\hspace{7mm}\makebox[10.7cm][s]{we have formalism locality.  The probability for a circuit is given by the}
\par
\noindent\hspace{7mm}\makebox[10.7cm][s]{corresponding duotensorial calculation (which is a scalar since there are no}
\par
\noindent\hspace{7mm}\makebox[10.7cm][s]{indices left over). We show how to put classical probability theory and}
\par
\noindent\hspace{7mm}\makebox[10.7cm][l]{quantum theory into this framework.}
}}


\section{Introduction}

Consider applying quantum theory to an arbitrary region of spacetime.  This region may be oddly shaped and even have disjoint parts.  In standard quantum theory we would need to use a larger region containing the given region but which had an initial and a final spacelike hypersurface. We could then evolve a quantum state from an initial state to a final state and use the machinery of quantum theory to make calculations pertaining to the given region.  In the histories formulation (Feynman's path integrals) we would consider histories over this larger region for the same purpose. In each case we need to consider mathematical objects pertaining to a larger region other than the region of interest.  This is inefficient.  Further, it reveals that we are not taking a truly spacetime approach.  A desirable property for a formulation to have is \cite{Flocality}
\begin{quote}
{\bf Formalism locality:} A formalism for a physical theory is said to have the property of \lq\lq formalism locality" if we can do calculations pertaining to any region of spacetime employing only mathematical objects associated with that region.
\end{quote}
Note that this is a property of the way a theory is formulated rather than being an intrinsic property of the physics itself.

In this paper we consider probabilistic theories that admit a circuit formulation.   This can be understood as an kind of operational formulation in which we have operations wired together.  An operation is represented diagrammatically by a box:
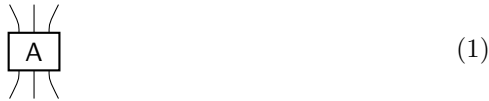
\vspace{-12pt}
\begin{equation}
\begin{Diagram}{0}{-0.1}
\Opbox{A}{0,0}
\inwire[-5]{A}{1}\inwire{A}{2}\inwire[5]{A}{3}
\outwire[-5]{A}{1}\outwire{A}{2}\outwire[5]{A}{3}
\end{Diagram}
\end{equation}
Each operation has inputs and outputs as shown (the inputs are at the bottom and the outputs are at the top). These are apertures through which we can imagine that systems (such as electrons and photons) travel. Each operation also has a setting (selected by positioning knobs for example) and an outcome set (read off meters or flashing lights for example).  Wires correspond to aligning apertures such that we can imagine systems passing between the operations.  A circuit is a bunch of operations wired together having no open inputs or outputs left over.  A fragment is a part of a circuit and can have open inputs and outputs.  Fragments are the circuit equivalent of arbitrary regions of spacetime.   We will develop a symbolic and diagrammatic notation for operational descriptions.

We will make two assumptions.  We will give more careful statements of these later once we have developed the necessary terminology.  Roughly speaking these assumptions are: (1) we can associate a probability with any circuit and this depends only on the properties of that circuit; (2) operations are fully decomposable in that they can be thought of as a linear combination of operations that consist of an effect for each input and a preparation for each output.   We can choose these effects and preparations from fiducial sets.  A choice of fiducial set constitutes, effectively, a choice of basis. If we do this, the coefficients in this linear combination give us one form of what we will call the {\it duotensor} associated with the given operation, namely
\begin{equation}
\begin{Diagram}{0}{0}
\Duobox{A}{0,0}
\inwhite[-5]{A}{1}\inwhite{A}{2}\inwhite[5]{A}{3}
\outwhite[-5]{A}{1}\outwhite{A}{2}\outwhite[5]{A}{3}
\end{Diagram}
\end{equation}
having all white dots.  We can change white dots to black dots on a duotensor by using the {\it hopping metric} $\bbdots$ (to be defined later).  A general duotensor can have black and white dots
\begin{equation}
\begin{Diagram}{0}{0}
\Duobox{A}{0,0}
\inblack[-5]{A}{1}\inwhite{A}{2}\inwhite[5]{A}{3}
\outblack[-5]{A}{1}\outwhite{A}{2}\outblack[5]{A}{3}
\end{Diagram}
\end{equation}
The entries in the duotensor having all black dots
\begin{equation}
\begin{Diagram}{0}{0}
\Duobox{A}{0,0}
\inblack[-5]{A}{1}\inblack{A}{2}\inblack[5]{A}{3}
\outblack[-5]{A}{1}\outblack{A}{2}\outblack[5]{A}{3}
\end{Diagram}
\end{equation}
are equal to the probabilities obtained by having fiducial preparations on all the inputs and fiducial effects on all the outputs (we call these the {\it fiducial probabilities}). The duotensor for a fragment is given by linking together the duotensors for the operations composing it in accordance with the fragments wiring.  For example,
\vspace{-14pt}
\begin{equation}\label{fraghasduo}
\text{the fragment}~~~~~
\begin{Diagram}{0}{-1.1}
\Opbox{A}{0,0} \Opbox[2]{B}{3,4} \Opbox[2]{C}{1,8}
\wire{A}{B}{3}{1}\wire{B}{C}{1}{2}
\inwire[-5]{A}{1}\inwire{A}{2}\inwire[5]{A}{3}
\outwire[-5]{A}{1}\outwire{A}{2}
\inwire[5]{B}{2}\outwire[5]{B}{2}
\inwire[-5]{C}{1} \outwire{C}{1.5}
\end{Diagram}
~~~~~\text{has duotensor}~~~~~
\begin{Diagram}{0}{0.2}
\Duobox{A}{0,0} \Duobox[2]{B}{4,-3} \Duobox[2]{C}{8,-1}
\link{A}{B}{3}{1}\link{B}{C}{1}{2}
\inblack[-5]{A}{1}\inwhite{A}{2}\inwhite[5]{A}{3}
\outblack[-5]{A}{1}\outwhite{A}{2}
\inblack[5]{B}{2}\outwhite[5]{B}{2}
\inwhite[-5]{C}{1} \outblack{C}{1.5}
\end{Diagram}
\end{equation}
When we link together duotensors like this we must join black and white dots.  Linking together black and white dots represents summing over the corresponding indices.  We can always turn a white dot into a black one using the hopping metric so we can link together duotensors whatever form they are presented to us in.  We can change the colour of the dots on the open inputs and outputs of the duotensor in (\ref{fraghasduo}) using the hopping metric (and the resulting object can still be understood as being the duotensor associated with the fragment in (\ref{fraghasduo})).

Duotensors are like tensors but have a bit more structure having to do with the fact that we effectively have two bases associated with each index - one for the effects and one for the preparations.  The black and white dots reflect this (as we will see later). We will develop a symbolic and diagrammatic notation for the mathematics associated with these duotensors.  The diagrammatic notation is basically the same as Penrose's diagramatic notation for tensors \cite{Penrosenotation, roadtoreality} (though he does not have black and white dots of course).  It is important to note that the diagrammatic notation is just as good as the symbolic notation for representing calculations involving duotensors and much more appealing visually.  As pointed out by Penrose \cite{roadtoreality}, one serious disadvantage of diagrammatic notation is typesetting the corresponding diagrams in published papers.  To ease this difficulty in the case of dutensors,  the \verb+duotenzor+ package was developed for drawing the necessary diagrams in LaTeX documents.  This allows diagrams to be entered in LaTeX much the same way symbolic mathematics is entered (though with about five times the effort).  See Appendix A for details.  All diagrams in this paper were drawn using this package.

We find that there is a striking correspondence between the notation for the operational description for a circuit or fragment and the notation for the corresponding mathematical calculation.  This is true both for symbolic and diagrammatic notation. The physical description and the associated mathematics look the same. In going from the operational description to the mathematical calculation we pass through hybrid diagrams which contain both operational and mathematical elements.  Assumption (2) can be presented by such a hybrid diagram (see (\ref{assumption2diagram}) in Sec.\ \ref{assumption2section}).

Probabilities associated with circuits are given by the corresponding duotensor calculation.  Since circuits have no open inputs or outputs, all indices are summed over in the duotensorial calculation and consequently we get a scalar (which is equal to the probability for the circuit).  We need to be a little more careful when considering probabilities for fragments.  In most cases we will not be able to associate a probability with a fragment since it is possible that outside influences can effect the fragment's outcomes (this is not the case with circuits).  We therefore have to take a
two-step approach.  First we give a condition under which such probabilities are independent of outside influences. Second, in the case that this condition is satisfied, we give an expression for the probability.  By adopting this two step approach we are able to have formalism locality.

Classical probability theory and quantum theory (for finite systems) can be uploaded into this framework.  To do this we must make a choice of fiducial effects and preparations.  Then we identify the elements of the duotensor having all black dots with the corresponding object in these theories (namely the fiducial probabilities).

The work in this paper is in the convex probabilities tradition.  It is also much influenced by the research of Abramsky, Coecke and co-workers \cite{AbramskyCoecke}.  In particular, we join the {\it quantum picturalism} revolution \cite{quantumpicturalism} initiated by these authors by making copious use of diagrams for doing calculations.  A review of the relevant literature is given in Sec.\ \ref{relatedwork}.

\section{Operational descriptions}

\subsection{Operations}

We will consider experiments that can be represented operationally by wiring together operations.
\begin{description}
\item[Operations:] An operation, $\mathsf{A}$, corresponds to one use of an apparatus and has the following features.
\begin{description}
\item[\it Inputs and outputs:] These are apertures we imagine a system can pass into and out of.  They come in various {\it types} $\mathsf a$, $\mathsf b$, \dots (which we can think of as corresponding to the type of system passing through the aperture).
\item[\it A Setting:] The setting, $\mathsf{s(A)}$, specifies the positions of the knobs and any other adjustables on the apparatus. The setting is part of the specification of $\mathsf A$.
\item[\it An outcome set:] Associated with any operation is an outcome set, $\mathsf{o(A)}$. The outcome, $\mathsf{x_A}$, could read off from a meter, from flashing lights, or from hearing whether a detector clicks for example.  If $\mathsf{x_A\in o(A)}$ then we say operation $\mathsf{A}$ \lq\lq happened".  The outcome set is part of the specification of $\mathsf A$.
\end{description}
\end{description}
An operation is represented graphically by a box with inputs and outputs.  Symbolically it is represented by displaying the inputs as subscripts and the outputs as superscripts
\begin{equation}\label{operationA}
\begin{Diagram}{0}{-0.80}
\Opbox{A}{3,3}
\inwire[-5]{A}{1}\Opsymbol{a}
\inwire{A}{2}\Opsymbol{b}
\inwire[5]{A}{3}\Opsymbol{b}
\outwire[-5]{A}{1.5}\Opsymbol{b}
\outwire[5]{A}{2.5}\Opsymbol{c}
\end{Diagram}
~~~~~~~~~~~~\Longleftrightarrow ~~~~~~~~~~~~~\mathsf{A_{a_1b_2b_3}^{b_4c_5}}
\end{equation}
The integers 1, 2, 3, \dots label the inputs and outputs.  These have no physical significance and we can relable them. However, the letters, $\mathsf{a}$, $\mathsf{b}$ are physically significant since they denote the types.

Each operation naturally belongs to a set of operations whose members correspond to one use of the same apparatus with the same setting but having different outcome sets $\mathsf {o_i(A)}$. It will sometimes be useful to include the specification of the outcome set in the notation for the operation.  In this case we will use the notation $\mathsf{ A[i]}$.

\subsection{Wires}\label{sectionwires}

The apertures on apparatuses can be aligned with one another.  We represent this by having wires go between operations.
\begin{description}
\item[Wires:] Outputs can be connected to inputs by wires.  For any collection of operations connected by wires we have the following wiring rules.
\begin{description}
\item[\it One wire:] At most one wire can be connected to any given input or output.
\item[\it Type matching:] Wires can connect inputs and outputs of the same type.
\item[\it No closed loops:]  Wires are directed (they go from output to input). We demand that if we trace forward along wires through the operations then we cannot get back to the same operation.  This corresponds to ruling out closed time-like loops.
\end{description}
\end{description}
Symbolically we will represent a wire by a repeated index.  For example,
\begin{equation}\label{ABwired}
\begin{Diagram}{0}{-1.2}
\Opbox{A}{7,0}
\Opbox[2]{B}{3,7}
\inwire[-5]{A}{1}\Opsymbol{a}
\inwire{A}{2}\Opsymbol{a}
\inwire[5]{A}{3}\Opsymbol{b}
\wire{A}{B}{1.5}{2}\opsymbol{b}
\outwire[5]{A}{2.5}\Opsymbol{c}
\inwire[-5]{B}{1}\Opsymbol{a}
\outwire[-5]{B}{1}\Opsymbol{d}
\outwire[5]{B}{2}\Opsymbol{c}
\end{Diagram}
~~~~~~~~ \Longleftrightarrow ~~~~~~~~~ \mathsf{A_{a_1a_2b_3}^{b_4c_5}}\mathsf{B_{a_5b_4}^{d_6c_7}}
\end{equation}
Note that the order of the symbols $\mathsf A$ and $\mathsf B$ is not significant since the causal ordering of the operations is indicated by the wires.  Thus,
\begin{equation}
\mathsf{A_{a_1a_2b_3}^{b_4c_5}}\mathsf{B_{a_5b_4}^{d_6c_7}} = \mathsf{\mathsf{B_{a_5b_4}^{d_6c_7}A_{a_1a_2b_3}^{b_4c_5}}}
\end{equation}
Likewise, in the diagrammatic notation, the placement of the boxes on the page and the lengths and shapes of the wires are not significant. All that matters is which output is wired into which input (i.e. the graphical information in the diagram is what is important).  In particular, there is no significance to the vertical position on the page - it does not calibrate a background Newtonian time.  This being said, it makes sense to arrange the boxes so that the wires generally go up the page otherwise the diagram becomes untidy.  The orientation of the boxes is significant.  The inputs go into the bottom of boxes, and the outputs come out the top.

There may be additional wiring rules.  For example, it may be impossible to place apparatuses together in certain ways because of their physical shape.  In fact, since an experiment may be dynamic, we are also aligning things in time and so it may be impossible to place apparatuses together in spacetime.

\subsection{Fragments}\label{fragments}

The object in (\ref{ABwired}) above is an example of a fragment.
\begin{description}
\item[Fragments:] A fragment is formed by wiring together a bunch of operations.  We will denote fragments by uppercase sans serif $\mathsf A$, $\mathsf B$, etc. (just as for operations which are, in fact, special cases of fragments).
\item[Example:]  For example, let the fragment $\mathsf E$ be given by
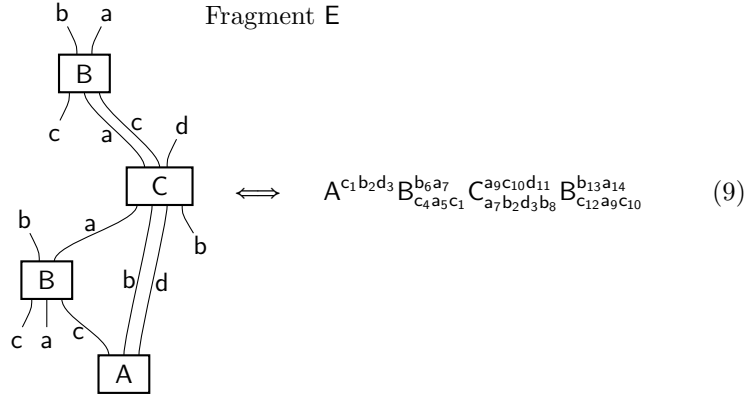
\begin{equation}\label{fragmentE}
\begin{Diagram}{-4}{-1}
\Opbox{A}{4.1,-6}\Opbox{B}{0,-1}\Opbox[4]{C}{6,4}\opbox{F}{2,10}\opsymbol{B}
\inwire[-5]{B}{1}\Opsymbol{c}\inwire{B}{2}\Opsymbol{a}\wire{A}{B}{1}{3}\opsymbol{c}\wire{A}{C}{2}{2}\opsymbol{b}\wire{A}{C}{3}{3}\otherside\opsymbol{d}
\outwire[-5]{B}{1.5}\Opsymbol{b}\wire{B}{C}{2.5}{1}\opsymbol{a}
\inwire[6]{C}{4}\Opsymbol{b}
\wire{C}{F}{1.5}{2}\opsymbol{a}\wire{C}{F}{2.5}{3}\otherside\opsymbol{c}\outwire[5]{C}{3}\Opsymbol{d}
\inwire[-5]{F}{1}\Opsymbol{c}
\outwire[-5]{F}{1.5}\Opsymbol{b}\outwire[5]{F}{2.5}\Opsymbol{a}
\placelatex{12,13}{\text{Fragment} \ensuremath{\mathsf E}}
\end{Diagram} \hspace{-1.6cm}
\Longleftrightarrow ~~~~ \mathsf{ A^{c_1b_2d_3} B_{c_4a_5c_1}^{b_6a_7} C_{a_7b_2d_3b_8}^{a_9c_{10}d_{11}} B_{c_{12} a_9 c_{10}}^{b_{13}a_{14}}}
\end{equation}
Notice that the operation $\mathsf B$ is used twice here (this corresponds to two separate uses of the same type of apparatus).
\item[Features:] A fragment will, in general, have some open inputs and outputs left over.  In particular, it may have outputs which could, in principle, be wired into inputs on the same fragment without violating the no closed loops assumption. It can be a part of a much bigger fragment.  {\it A fragment is therefore the circuit language equivalent of an arbitrary region of space-time.}  We allow a fragment to consist of disjoint parts (the circuit equivalent of an arbitrary region of space-time consisting of disjoint parts).  The outcome $\mathsf{ x_E}$ is given by specifying the outcome at each operation making up the fragment.  In specifying the fragment, $\mathsf E$, we give an outcome set $\mathsf{ o(E)}$, the settings $\mathsf {s(E)}$ (specifying this means specifying a setting at each operation), and the wiring $\mathsf{w(E)}$.  We will usually denote the settings and wiring by $\mathsf{sw(E)}$ for brevity.  Note that we are free to have outcome sets $\mathsf{o(E)}$ that cannot be written as the cartesian product of outcome sets at each of the operations making up the circuit fragment.  The statement that a given fragment \lq\lq has happened" means that the experiment with all the corresponding apparatuses placed together in accordance with the given settings and wiring, $\mathsf{sw(E)}$, was performed and the outcome, $\mathsf{x_E}$, was in the given outcome set, $\mathsf{o(E)}$.
\item[Experiments:]  Each fragment naturally belongs to a class of fragments corresponding to the same apparatuses with the same knob settings and the same wiring but having different outcome sets.  We will say that these fragments correspond to \lq\lq the same experiment".  We will denote the members of such a set of fragments as $\mathsf{ E[i]}$, the outcome sets by $\mathsf{ o_i(E)}$, and the settings and wirings by $\mathsf{ sw_i(E)}$.
\end{description}

\subsection{Circuits}

\begin{description}
\item[Circuits:] A circuit is a special case of a fragment in which there are no open inputs or outputs left over. Since they are special cases of fragments we will denote them by uppercase sans serif font, $\mathsf A$, $\mathsf B$, \dots.  For example, let the circuit $\mathsf H$ be
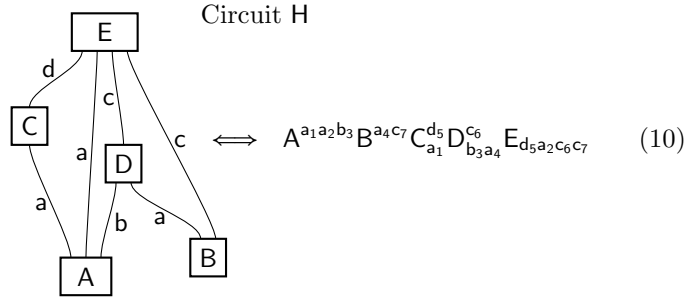
\begin{equation}
\begin{Diagram}{0}{-1.4}
\Opbox{A}{8,-2}\Opbox[2]{B}{14.5,-1}\Opbox[2]{C}{5,6}\Opbox[2]{D}{10,4}\Opbox[4]{E}{9,11}
\wire{A}{C}{1}{1.5}\opsymbol{a}\wire{A}{E}{2}{2}\opsymbol{a}\wire{A}{D}{3}{1}\otherside\opsymbol[5,0]{b}
\wire{B}{D}{1}{2}\opsymbol{a}\wire{B}{E}{2}{4}\otherside\opsymbol{c}\wire{D}{E}{1.5}{3}\opsymbol{c}\wire{C}{E}{1.5}{1}\opsymbol[0,8]{d}
\placelatex{17,12}{\text{Circuit} \ensuremath{\mathsf H} }
\end{Diagram}
\hspace{-1.5cm}
\Longleftrightarrow ~~  \mathsf{   A^{a_1a_2b_3} B^{a_4c_7} C_{a_1}^{d_5} D_{b_3a_4}^{c_6} E_{d_5a_2c_6c_7}  }
\end{equation}
Circuits can consist of disjoint parts.
\item[Experiments] Each circuit naturally belongs to a set of circuits corresponding to the same apparatuses and having the same wiring and knob settings but different outcome sets at the operations.  We will denote the members of such a set of circuits by $\mathsf{H[i]}$.
\end{description}

\subsection{Closing inputs and outputs}

It is useful to add standard operations that allow us to {\it close} inputs and outputs on operations. Closing an input on an operation corresponds to placing some standard device for that type in front of that aperture (it may, for example, input an electron prepared in the spin up direction).  Closing an output corresponds to simply placing a block after that aperture that absorbs the system.  We can, for example, close some inputs and outpus on (\ref{ABwired})
\begin{equation}
\begin{Diagram}{0}{-1.2}
\Opbox{A}{7,0}
\Opbox[2]{B}{3,7}
\inwire[-5]{A}{1}\Opsymbol{a}
\closedinwire{A}{2}\Opsymbol{a}
\inwire[5]{A}{3}\Opsymbol{b}
\wire{A}{B}{1.5}{2}\opsymbol{b}
\closedoutwire[5]{A}{2.5}\Opsymbol{c}
\closedinwire[-5]{B}{1}\Opsymbol{a}
\outwire[-5]{B}{1}\Opsymbol{d}
\closedoutwire[5]{B}{2}\Opsymbol{c}
\end{Diagram}
~~~~~~~~ \Longleftrightarrow ~~~~~~~~~ \mathsf{A_{a_1\underline{a}_2b_3}^{b_4\overline{c}_5}}\mathsf{B_{\underline{a}_5b_4}^{d_6\overline{\mathsf c}_7}}
\end{equation}
We have indicated closing an input or output diagrammatically by a \ultrathickdash at the end of the wire and symbolically by underlining inputs and overlining outputs.  We can turn any fragment into a circuit by closing all the inputs and outputs.

\subsection{Notation}

We use sans-serif font for the operational description (we do this since we are saving the normal maths font for the corresponding duotensors to be introduced later).  We use upper case sans-serif letters $\mathsf{A}$, $\mathsf B$, \dots to represent operations, circuit fragments, and circuits (circuits and operations are special cases of fragments in any case).  We use lower case sans-serif letters $\mathsf{a}$, $\mathsf b$ to represent types. We can group together two types and regard them as a single type (this corresponds to a composite system).   Composite types, such as $\mathsf{ ab}$ may be represented by a single letter $\mathsf d$.

The diagrammatic and symbolic representations we have given encode the three wiring constraints given above in Sec.\ \ref{sectionwires}. However, it is not rich enough as it stands to encode any additional wiring constraints.  If there are any such additional constraints and they play an important role then we would need to develop richer diagrammatic and symbolic representations.

Fragment $\mathsf E$ in (\ref{fragmentE}) above has four inputs and four outputs left over.  For some purposes it will be useful to represent this by the abbreviated notation
\begin{equation}
\mathsf{  E_{c_1}^{b_6}{}_{a_5}^{d_{11}}\!{}_{b_{12}}^{b_{13}}{}_{c_{12}}^{a_{14}}   }
\end{equation}
This does not, however, convey information about the causal structure  - namely which output can lead to which input.  If we wish to convey that information we must use a richer notation such as in (\ref{fragmentE}).

\section{Probabilities}

\subsection{Probabilities for fragments}

A fragment, $\mathsf A$, is something that happens.  We can consider the probability that a fragment happens. This is the probability that we see an outcome $\mathsf{x_A}$ in $\mathsf{o(A)}$ given that we have knob settings $\mathsf{s(A)}$ and wiring $\mathsf {w(A)}$. More generally, we consider probabilities such as
\begin{equation}
\text{Prob}\mathsf{(A|B)}
\end{equation}
This is shorthand notation for the probability
\begin{equation}
\text{Prob}\mathsf{\big(x_A\in o(A) | sw(A), sw(B), x_B\in o(B)\big)}
\end{equation}
In such expressions we will always take the fragments, $\mathsf A$, $\mathsf B$, $\mathsf C$, \dots to be non-overlapping.  By non-overlapping we mean that they have no operations in common (though, of course, they can have different instances of the same operation).  Non-overlapping fragments may be connected by wires.  We will include subscripts and superscripts when they help to clarify the situation (see example for spin measurements in Sec.\ \ref{wcprobs} below).  They may also be connected to other fragments not listed.

\subsection{Well conditioned probabilities}\label{wcprobs}

Now we introduce an important definition.
\begin{quote}
{\bf Well conditioned probabilities:} We will say that the probability $\text{Prob}\mathsf{(A|B)}$ is {\it well conditioned} if
\begin{equation}
\text{Prob}\mathsf{(A|BC)} = \text{Prob}\mathsf{(A|BD)}  ~~\text{for all} ~~ \mathsf{C} ~\text{and}~ \mathsf{D}
\end{equation}
in which case $\text{Prob}\mathsf{(A|B)}$ is fully determined by $\mathsf{A}$ and $\mathsf{B}$.
\end{quote}
In most cases,  probabilities such as $\text{Prob}\mathsf{(A|B)}$ will not be well conditioned.  Consider an experiment from quantum physics.  Imagine we we have a device $\mathsf{ A^{a_1}}$ which prepares a spin half particle (which we take to be of type $\mathsf a$) in the up state followed, in sequence, by two spin measurements $\mathsf{ B_{a_1}^{a_2}}$ and  $\mathsf{ C_{a_2}^{a_3}}$, along some directions, and then followed by an operation $\mathsf{ D_{a_3}^{b_4}}$ which may be a spin measurement or something else.
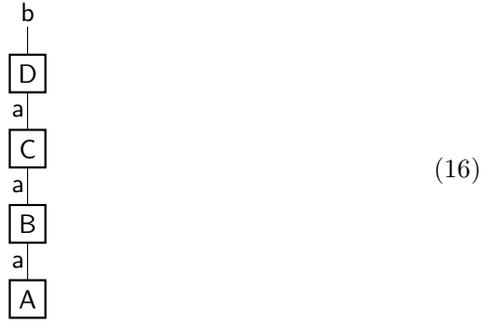
\vspace{-10pt}
\begin{equation}
\begin{Diagram}{0}{-1.8}
\Opbox[2]{A}{0,0}
\Opbox[2]{B}{0,4}
\Opbox[2]{C}{0,8}
\Opbox[2]{D}{0,12}
\wire{A}{B}{1.5}{1.5}\opsymbol{a}
\wire{B}{C}{1.5}{1.5}\opsymbol{a}
\wire{C}{D}{1.5}{1.5}\opsymbol{a}
\outwire{D}{1.5}\Opsymbol{b}
\end{Diagram}
\end{equation}
The following three examples illustrate the notion of a well conditioned probability.
\begin{enumerate}
\item The probability $\text{Prob}\mathsf{(C_{a_2}^{a_3}|A^{a_1})}$ is clearly not well conditioned since it depends on which direction the spin is measured along at $\mathsf B$, i.e.\
\begin{equation}
\text{Prob}\mathsf{(C_{a_2}^{a_3}|A^{a_1}B_{a_1}^{a_2})} \not=\text{Prob}\mathsf{(C_{a_2}^{a_3}|A^{a_1}\tilde{B}_{a_1}^{a_2})}
\end{equation}
where $\mathsf{\tilde{B}_{a_1}^{a_2}}$ is a spin measurement along a different direction from $\mathsf{B_{a_1}^{a_2}}$.
\item Perhaps a little more surprisingly, the probability $\text{Prob}\mathsf{(B_{a_1}^{a_2}|A^{a_1})}$ is also not well conditioned because
\begin{equation}
\text{Prob}\mathsf{(B_{a_1}^{a_2}|A^{a_1}C_{a_2}^{a_3})} \not=\text{Prob}\mathsf{(B_{a_1}^{a_2}|A^{a_1}\tilde{C}_{a_2}^{a_3})}
\end{equation}
(where $\mathsf{\tilde{C}_{a_2}^{a_3}}$ is a spin measurement along a different direction to $\mathsf{C_{a_2}^{a_3}}$) because postselection effects the probability (as a simple calculation will show).  To get a well conditioned probability for $\mathsf B$ given $\mathsf A$ in this situation we could follow the $B$ operation by a closing operation. I.e.\ the $\text{Prob}\mathsf{(B_{a_1}^{\overline{a}_2}|A^{a_1})}$ is well conditioned.
\item The pre- and post-selected probability $\text{Prob}\mathsf{(B_{a_1}^{a_2}|A^{a_1}C_{a_2}^{a_3})}$ is well conditioned since
\begin{equation}
\text{Prob}\mathsf{(B_{a_1}^{a_2}|A^{a_1}C_{a_2}^{a_3}D_{a_3}^{a_4})}=\text{Prob}\mathsf{(B_{a_1}^{a_2}|A^{a_1}C_{a_2}^{a_3}\tilde{D}_{a_3}^{a^4})} \end{equation}
for any $\mathsf{D_{a_3}^{b_4}}$ and $\mathsf{\tilde{D}_{a_3}^{b_4}}$.  This is true because $\mathsf C$ is a complete spin measurement (corresponding to a non-degenerate observable) and so subsequent postselection does not effect this probability.
\end{enumerate}

\subsection{Assumption 1}

We will make two assumptions to set up the framework in this paper.  We state the first here.
\begin{quote}
{\bf Assumption 1} The probability, $\text{Prob}(\mathsf A)$, for any circuit, $\mathsf A$ (this has no open inputs or outputs), is well conditioned.  It is therefore determined by the operations and the wiring of the circuit alone and is independent of settings and outcomes elsewhere.
\end{quote}
In fact this can be regarded, in part, as being as an assumption about how well we specify the properties of the sets of operations we consider.  If we consider a set of operations for which Assumption 1 is true then we can form a set of operations for which it is not true in the following way.  We take some of the operations in the set and delete some inputs and some outputs from our specification of them.  To give a quantum example, we may have some two qubit gates yet cross out one of the inputs and one of the outputs from our specification.  Then a circuit which contained one of these and appeared to be closed would actually be open (with respect to the original specification of the operations).  Using the deleted inputs and outputs, it would be possible to send in a qubit from an operation that was not part of the circuit (according to the new specification) and hence Assumption 1 would not be true.  Assumption 1 asserts, in part, that all inputs and outputs have been correctly identified and specified.

It follows from Assumption 1 that if $\mathsf A$ and $\mathsf B$ are both circuits then, for the composite circuit $\mathsf{ AB}$ (consisting of the disconnected parts $\mathsf A$ and $\mathsf B$), we have
\begin{equation}\label{probfatorises}
\text{Prob}(\mathsf{AB})=\text{Prob}(\mathsf{A})\text{Prob}(\mathsf{ B})
\end{equation}
because
\begin{flalign*}
\text{Prob}&(\mathsf{AB}) = \text{Prob}\big(\mathsf{x_A\in o(A), x_B\in o(B)|\,sw(A), sw(B)}\big)  \\
 =& \text{Prob}\big(\mathsf{x_A\in o(A)| x_B\in o(B), sw(A), sw(B)}\big) \text{Prob}\big(\mathsf{x_B\in o(B)|\, sw(A), sw(B)}\big)  \\
 =& \text{Prob}\big(\mathsf{x_A\in o(A)|\, sw(A)}\big) \text{Prob}\big(\mathsf{ x_B\in o(B)|\, sw(B)}\big)
\end{flalign*}
where we use Bayes rule in the second line and Assumption 1 in the third line.

\subsection{Well conditioned probability ratio}\label{probabilityratio}

We introduce a further definition.
\begin{quote}
{\bf Well conditioned probability ratio:} We will say the probability ratio
\begin{equation}\label{probratio}
\frac{\text{Prob}\mathsf{(A[i])}}{\text{Prob}\mathsf{(A[j])}}
\end{equation}
is well conditioned if
\[ \frac{\text{Prob}\mathsf{(A[i]\,|\,C)}}{\text{Prob}\mathsf{(A[j]\,|\,C)}}  \]
is independent of $\mathsf C$ where $\mathsf C$ is any fragment which completes $\mathsf A[i]$ (and $\mathsf A[j]$) into a circuit.   Here $\mathsf A[i]$ and $\mathsf A[j]$ correspond to different outcome sets for the same experiment.  By Bayes rule, this equivalent to demanding that
\[ \frac{\text{Prob}\mathsf{(A[i]C)}}{\text{Prob}\mathsf{(A[j]C)}} \]
is independent of $\mathsf C$.  According to Assumption 1 both the numerator and denominator of this expression are well defined so we this is a test we can run.
\end{quote}
If $\text{Prob}\mathsf{(A[i])}$ and $\text{Prob}\mathsf{(A[j])}$ are each well conditioned then it follows that their ratio is.  However, it is possible that, taken separately, they are not well conditioned but that the ratio is.  If we define probabilities as long-run relative frequencies then the probability ratio is equal to the number of times $\mathsf{A[i]}$ happens divided by the number of times $\mathsf{A[j]}$ happpens in the long run.

A special case is where the outcome set $\mathsf{o_j}$ for the fragment $\mathsf{A[j]}$ is the set of all possible outcomes. We denote this $\mathsf{A[I]}$. Since some outcome must happen we know that $\text{Prob}\mathsf{(A[I]\,|\, B)}=1$ for any $\mathsf B$.  We can, then, regard the probability $\text{Prob}(\mathsf{ A[i]\,|\,B })$ as a probability ratio (using Bayes rule)
\begin{equation}\label{ratiotocondition}
\text{Prob}\mathsf{(A[i]\,|\,B)}=\frac{\text{Prob}\mathsf{(A[i]B)}}{\text{Prob}\mathsf{(A[I]B)}}
\end{equation}
Hence, the idea of a probability ratio is more general than that of a conditional probability.

\subsection{Objective}\label{ourobjective}

A probability ratio that is not well conditioned is not well defined.  Whatever value we write down for it could be made to be wrong by an adversary who has control over other conditions that would effect the outcome we are looking at.  Therefore, we cannot expect a physical theory to predict the values of probability ratios that are not well conditioned.  However, it is reasonable to expect our physical theory to tell us whether a probability ratio is well conditioned.
\begin{description}
\item{\bf Our objective} is to construct a mathematical framework for theories which
\begin{enumerate}
\item Associates mathematical objects with all fragments.
\item Provides a mathematical condition for saying whether the probability ratio
\[\frac{\text{Prob}\mathsf{(A[i])}}{\text{Prob}\mathsf{(A[j])}}   \]
for any $\mathsf{A[i]}$ and $\mathsf{A[j]}$ (for the same experiment) is well conditioned employing only mathematical objects associated with these fragments.
\item In the case that the probability ratio is well conditioned, provides an expression saying what it is equal to employing only mathematical objects associated with the fragments $\mathsf{A[i]}$ and $\mathsf{A[j]}$.
\end{enumerate}
\end{description}
If we can do this we will have the formalism locality property.

This is not the approach that is taken in most existing formulations of physical theories.  Rather, we usually have equations that apply only to specially shaped spacetime regions for which the probabilities are necessarily well conditioned.  For example, in theories which evolve a state, these specially shaped regions must have an initial and a final space-like hypersurface.  To make statements about arbitrary spacetime regions in such theories, we need to first apply the theory to the specially shaped regions.  This is the reason that these formulations do not have the formalism locality property.

It would be impossible to identify special shaped regions if we had indefinite causal structure  as we expect in a theory of quantum gravity. In such a case the above approach seems like the most sensible choice.  This is not an immediate issue for the present paper since we are considering circuits whose wires define a causal structure.  Nevertheless, the approach in this paper was motivated by the problem of quantum gravity.  Further, quantum theory with well defined causal structure may ultimately be best understood as a limiting case of a more general theory of quantum gravity having indefinite causal structure.  In this case, it is likely that the formulation of quantum theory we would most naturally arrive at in such a limitting procedure would be one having the formalism locality property.

\section{Equivalence relations}\label{equivalencerelations}

We define the function $p(\cdot)$ as follows
\begin{equation}
p(\alpha \mathsf{A} + \beta \mathsf{B} + \dots) := \alpha\text{Prob}(\mathsf{A}) + \beta\text{Prob}(\mathsf{B}) +\dots
\end{equation}
for circuits $\mathsf A$, $\mathsf B$, \dots. and real numbers $\alpha$, $\beta$, \dots (these can be negative).  Clearly $p(\mathsf{A})=\text{Prob}(\mathsf{A})$.  Note that $p(\cdot)$ is defined for circuits and not for fragments with open inputs and/or outputs.  It follows from (\ref{probfatorises}) that
\begin{equation}\label{pfactorises}
p(\mathsf{AB})=p(\mathsf{A})p(\mathsf{ B})
\end{equation}

We will use the function $p(\cdot)$ to define equivalence relations.  To illustrate this, we will first look at an example.  Consider a  circuit composed of two fragments $\mathsf{A[i]\,_{a_1b_2}^{c_3d_4}}$ and $\mathsf{B_{c_3d_4}^{a_1b_2}}$.  The circuit can be written $\mathsf{A[i]\,_{a_1b_2}^{c_3d_4}B_{c_3d_4}^{a_1b_2}}$.  Assume, for the first three of the outcome sets labeled by $\mathsf i$, that $\mathsf{ o_1(A)}$ and $\mathsf{ o_2(A)}$ are disjoint and $\mathsf {o_3(A)=o_2(A)+o_1(A)}$.  Then we have
\begin{equation}
\text{Prob}(\mathsf{A[3]\,_{a_1b_2}^{c_3d_4}B_{c_3d_4}^{a_1b_2}})
= \text{Prob}(\mathsf{A[1]\,_{a_1b_2}^{c_3d_4}B_{c_3d_4}^{a_1b_2}})+ \text{Prob}(\mathsf{A[2]\,_{a_1b_2}^{c_3d_4}B_{c_3d_4}^{a_1b_2}})
\end{equation}
because probabilities are additive across disjoint outcome sets.  Hence
\begin{equation}
p(\mathsf{A[3]\,_{a_1b_2}^{c_3d_4}B_{c_3d_4}^{a_1b_2}})
= p\big(\mathsf{(A[1]\,_{a_1b_2}^{c_3d_4}+ A[2]\,_{a_1b_2}^{c_3d_4})B_{c_3d_4}^{a_1b_2}}\big)
\end{equation}
Given this we will write
\begin{equation}
\mathsf{A[3]\,_{a_1b_2}^{c_3d_4}
\equiv A[1]\,_{a_1b_2}^{c_3d_4}+ A[2]\,_{a_1b_2}^{c_3d_4}}
\end{equation}
We need to define what we mean by the use of the symbol $\equiv$.
\begin{quote}
{\bf Equivalence:}  We write
\begin{equation}\label{equivalence}
\text{expression}_1 \equiv \text{expression}_2
\end{equation}
(in words we will say that $\text{expression}_1$ is equivalent to $\text{expression}_2$)
if
\begin{equation}\label{defequiv}
p(\text{expression}_1 \mathsf{E}) \equiv p(\text{expression}_2 \mathsf{E})
\end{equation}
for any fragment $\mathsf E$ that makes the contents of the argument on both sides of this equation into a linear sum of circuits (as required for $p(\cdot)$ to be defined).
\end{quote}
There are two types of equivalence that satisfy this definition (Greek letters represent real numbers below):
\begin{enumerate}
\item Each expression is a real number plus a linear combination of circuits:
\begin{equation}
\alpha + \beta\mathsf{A} + \gamma \mathsf{B} + \dots \equiv \delta + \epsilon \mathsf{ C}+ \zeta \mathsf{D} + \dots
\end{equation}
where $\mathsf{A}$, $\mathsf B$, \dots, $\mathsf C$, $\mathsf D$, \dots, are all circuits.  In this case possible choices for $\mathsf E$ in (\ref{defequiv}) are any circuit.
\item Each expression is a linear combination of fragments
\begin{equation}
\alpha\mathsf{A} + \beta \mathsf{B} + \dots \equiv \gamma\mathsf{ C}+ \delta \mathsf{D} + \dots
\end{equation}
where $\mathsf{A}$, $\mathsf B$, \dots, $\mathsf C$, $\mathsf D$, \dots, are all fragments having the same causal structure (and therefore the same number and types of inputs and outputs). In this case possible choices for $\mathsf E$ in (\ref{defequiv}) are fragments that can be wired with any of the fragments $\mathsf{A}$, $\mathsf B$, \dots $\mathsf C$, $\mathsf D$, \dots to produce a circuit.
\end{enumerate}
An example of the first kind is  $\mathsf{ A}\equiv \frac{2}{3}$ for a circuit $\mathsf A$.  Given the way the $p(\cdot)$ function is defined, this would mean that $\text{Prob}(\mathsf {A})=\frac{2}{3}$.  In fact, in general, we have
\begin{equation}\label{AequivtoprobA}
\mathsf{ A} \equiv \text{Prob}(A) ~~~ \text{for any circuit} ~~ \mathsf{A}
\end{equation}
Equivalence is a weaker notion than equality.  Two things that are equivalent will, generally, be different things.  They are equivalent so far as the $p(\cdot)$ function is concerned.  Roughly speaking, two things are equivalent if they have the same probabilistic properties.  The notion of equivalence will prove to be very useful.  By making a second assumption we will be able to reduce complicated general circuits to a linear combination of a product of standard (fiducial) circuits.  This will enable us to obtain an expression for the probability for such general circuits.


\section{A simple circuit}\label{simplecircuit}

\subsection{Fiducial effects}

Consider an operation, $\mathsf{B_{a_1}}$, that has an input for a system of type $\mathsf{a}$ (we call such an operation, with no output, an {\it effect}).  We will say that $\mathsf{X}_\mathsf{a_1}^{a_1}$, where $a_1=1$ to $K_\mathsf{a}$, is a {\it fiducial set} of effects  if we can write
\begin{equation}\label{fideffects}
\mathsf{B}_\mathsf{a_1} \equiv B_{a_1} \mathsf{X}_\mathsf{a_1}^{a_1}
\end{equation}
for any effect $\mathsf{B}_\mathsf{a_1}$ and where the set is minimal in the sense that no set of fewer than $K_\mathsf{a}$ effects can have this property.  We are adopting the standard convention that we sum over the repeated indices ($a_1$ in this case).  Here $B_{a_1}$ supplies the coefficients.  There are $K_\mathsf{a}$ entries in $B_{a_1}$ which are real and can be negative.  We can always find a fiducial set like this since, in the worst case, we simply have one fiducial effect for every effect (that is equal to that effect).  However, in general, we expect the number of fiducial effects to be much smaller than the number of effects (this is certainly the case in finite dimensional quantum theory).  Note that equation (\ref{fideffects}) has mixed font.  Symbols in sans serif represent the operational description (the physics).  Symbols in standard font represent the mathematics.  Equation (\ref{fideffects}) is a hybrid equation since it has both types of font. It relates physics (the operational description) to mathematics!  We can write the elements of this equation in diagrammatic form.  Thus we put,
\begin{equation}
 \mathsf{X}_\mathsf{a_1}^{a_1}~~~~\Longleftrightarrow ~~
\begin{Diagram}{0}{0.1}
\fideffect{X}{0,0} \outblack{X}{1}\Duosymbol{a} \inwire{X}{1}\Opsymbol{a}
\end{Diagram}
\end{equation}
Note we introduce a black dot on this fiducial effect - this will be important.  This has mixed font.  The \lq\lq$\mathsf a$" indicates that the effect has a system of type $\mathsf a$ as its input.  The \lq\lq$a$" indicates that we have a index $a_1$ which runs over $a_1=1$ to $K_{\mathsf a}$. We could, of course, choose a different integer in the subscript to $a$.  For example, we could use the index $a_2$.  However, this integer is of no physical significance so we omit it in diagrammatic representations where it is not necessary.  We also put
\begin{equation}
B_{a_1} ~~~~\Longleftrightarrow ~~
\begin{Diagram}{0}{-0.1}
\Duobox[2]{B}{0,0}
\inwhite{B}{1.5}\Duosymbol{a}
\end{Diagram}
\end{equation}
Equation (\ref{fideffects}) can be written in both symbolic and diagrammatic form
\begin{equation}
\mathsf{B}_\mathsf{a_1}\equiv B_{a_1} \mathsf{X}_\mathsf{a_1}^{a_1}~~~~~\Longleftrightarrow  ~~~~~
\begin{Diagram}{0}{0}
\Opbox[2]{B}{0,0}\inwire{B}{1.5}\Opsymbol{a}
\end{Diagram}
~~\equiv~~
\begin{Diagram}{0}{0.07}
\fideffect{X}{0,-7pt}\Duobox[2]{B}{5,0}
\inwire{X}{1}\Opsymbol{a} \linkbw{X}{B}{1}{1.5} \duosymbol{a}
\end{Diagram}
\end{equation}
The horizontal link interupted by a black and a white dot indicate that we sum over the index corresponding to $a$.  In these diagrams we impose the rule that we must match black and white dots.   We define
\begin{equation}
\begin{Diagram}{0}{0.07}
\fideffect{X}{0,-7pt}\Duobox[2]{B}{5,0}
\inwire{X}{1}\Opsymbol{a} \linkbw{X}{B}{1}{1.5} \duosymbol{a}
\end{Diagram}
~~ := ~~
\begin{Diagram}{0}{0.07}
\fideffect{X}{0,-7pt}\Duobox[2]{B}{4,0}
\inwire{X}{1}\Opsymbol{a} \link{X}{B}{1}{1.5} \duosymbol{a}
\end{Diagram}
\end{equation}
This is unambiguous at this stage since the fiducial effect must have a black dot so the $B$ box must have a white dot.  This is a hybrid diagram.  Hybrid diagrams have wires running up for operational description and links running to the left for the mathematics. Horizontal links between boxes represent the summation over the corresponding index.

\subsection{Fiducial preparations}

Now consider an operation, $\mathsf{A^{a_1}}$, that outputs a system of type $\mathsf{a}$. We call operations having no inputs, such as this, {\it preparations}.  We will say that the set of preparations ${}_{a_1}\!\mathsf{X}^\mathsf{a_1}$, where $a_1=1$ to $K_\mathsf{a}$, constitutes a {\it fiducial set} of preparations for a system of type $\mathsf a$ if we can write
\begin{equation}\label{ABcircuit}
\begin{Diagram}{0}{-0.2}
\Opbox[2]{A}{0,-7pt} \outwire{A}{1.5} \Opsymbol{a}
\end{Diagram}
~~ \equiv ~~
\begin{Diagram}{0}{-0.2}
\Duobox[2]{A}{0,-7pt} \fidprep{X}{5,0} \linkwb{A}{X}{1.5}{1}\duosymbol{a} \outwire{X}{1}\Opsymbol{a}
\end{Diagram}
~~~\Longleftrightarrow ~~~ \mathsf{A^{a_1}} \equiv {}^{a_1}\!\!A \,\,\,{}_{a_1}\!\mathsf{X}^\mathsf{a_1}
\end{equation}
(we sum over $a_1$) for any preparation $\mathsf{A^{a_1}}$.  We represent the elements of this equation by
\begin{equation}
\begin{Diagram}{0}{-0.1}
\fidprep{X}{0,0} \inblack{X}{1} \Duosymbol{a} \outwire{X}{1} \opsymbol{a}
\end{Diagram}
~~\Longleftrightarrow ~~ {}_{a_1}\!\mathsf{X}^\mathsf{a_1} ~~~~~~~~~~~~~~~~
\begin{Diagram}{0}{-0.05}
\Duobox[2]{A}{0,0} \outwhite{A}{1.5}\duosymbol{a}
\end{Diagram}
\Longleftrightarrow ~~  {}^{a_1}\!\!A
\end{equation}
Note that, again, we have a black dot on the fiducial element and we match black and white dots (in accordance with rule introduced above). We define
\begin{equation}
\begin{Diagram}{0}{0}
\Duobox[2]{A}{0,-7pt} \fidprep{X}{5,0} \linkwb{A}{X}{1.5}{1}\duosymbol{a} \outwire{X}{1}\Opsymbol{a}
\end{Diagram}
~~ := ~~
\begin{Diagram}{0}{0}
\Duobox[2]{A}{0,-7pt} \fidprep{X}{4,0} \link{A}{X}{1.5}{1}\duosymbol{a} \outwire{X}{1}\Opsymbol{a}
\end{Diagram}
\end{equation}
Once again, this is not ambiguous at this stage since the fiducial preparation must have a black dot and so the $A$ box must have a white dot.

\subsection{The simple circuit}

Now consider the circuit:
\begin{equation}
\begin{Diagram}{0}{-0.6}
\Opbox[2]{A}{0,0}\Opbox[2]{B}{0,4} \wire{A}{B}{1.5}{1.5}\opsymbol{a}
\end{Diagram}
~~\equiv~~
\begin{Diagram}{0}{-0.6}
\Duobox[2]{A}{0.5,-7.5pt} \fidprep{X2}{5,0} \linkwb{A}{X2}{1.5}{1}\duosymbol{a}
\fideffect{X}{5,4}\Duobox[2]{B}{9.5,4cm+7pt}
\linkbw{X}{B}{1}{1.5} \duosymbol{a}
\wire{X2}{X}{1}{1}\opsymbol{a}
\end{Diagram}
~~=~~
\begin{Diagram}{0}{-0.6}
\Duobox[2]{A}{0.5,-7.5pt} \fidprep{X2}{4,0} \link{A}{X2}{1.5}{1}\duosymbol{a}
\fideffect{X}{4,4}\Duobox[2]{B}{7.5,4cm+7pt}
\link{X}{B}{1}{1.5} \duosymbol{a}
\wire{X2}{X}{1}{1}\opsymbol{a}
\end{Diagram}
\end{equation}
In symbolic form this is
\begin{equation}
\mathsf{A^{a_1}B_{a_1}} \equiv {}^{a'_1\,}\!\!A\,\,\, {}_{a'_1\!}\!\mathsf{X}^\mathsf{a_1} \,\,\,
\mathsf{X}_\mathsf{a_1}^{a_1}\,\,\,  B_{a_1}
\end{equation}
Using the linearity of the $p(\cdot)$ function we have
\begin{equation}\label{ABdiagramcalculation}
\begin{Diagram}{0}{-0.6}
\Duobox[2]{A}{0.5,-7.5pt} \fidprep{X2}{5,0} \linkwb{A}{X2}{1.5}{1}\duosymbol{a}
\fideffect{X}{5,4}\Duobox[2]{B}{9.5,4cm+7pt}
\linkbw{X}{B}{1}{1.5} \duosymbol{a}
\wire{X2}{X}{1}{1}\opsymbol{a}
\end{Diagram}
~\equiv ~
\begin{Diagram}{0}{0}
\Duobox[2]{A}{0,0} \bbinsert[0.8]{g}{4,0} \duosymbol{a} \Duobox[2]{B}{8,0}
\link{A}{g}{1.5}{1}\link{g}{B}{1}{1.5}
\end{Diagram}
\end{equation}
where we define
\begin{equation}\label{defbb}
\begin{Diagram}{0}{-0.1}
\bbmetric{g}{0,0}\duosymbol{a}
\end{Diagram}
~:=~
p\left(\!\!\!\!
\begin{Diagram}{0}{-0.6}
\fidprep{X2}{4,0}
\fideffect{X}{4,4}
\wire{X2}{X}{1}{1}\opsymbol{a}
\inblack{X2}{1} \Duosymbol{a}
\outblack{X}{1}\Duosymbol{a}
\end{Diagram}
\!\!\!\!\right)
\end{equation}
Note, it follows from this definition (using (\ref{AequivtoprobA})) that
\begin{equation}\label{defbbequiv}
\begin{Diagram}{0}{-0.6}
\fidprep{X2}{4,0}
\fideffect{X}{4,4}
\wire{X2}{X}{1}{1}\opsymbol{a}
\inblack{X2}{1} \Duosymbol{a}
\outblack{X}{1}\Duosymbol{a}
\end{Diagram}
~\equiv ~
\begin{Diagram}{0}{-0.1}
\bbmetric{g}{0,0}\duosymbol{a}
\end{Diagram}
\end{equation}
We will call $\bbdots$ the {\it hopping metric} (for reasons that will be clear below).  We also define
\begin{equation}\label{ABhoppingdiagram}
\begin{Diagram}{0}{-0.1}
\Duobox[2]{A}{0,0} \outblack{A}{1.5}
\end{Diagram}
\!\!\!\!:=~
\begin{Diagram}{0}{-0.1}
\Duobox[2]{A}{0,0} \outwhite{A}{1.5} \bbmetric[0.7]{g}{3.7,0}
\end{Diagram}
~~~~~~~~~~
\begin{Diagram}{0}{-0.1}
\Duobox[2]{B}{0,0} \inblack{A}{1.5}
\end{Diagram}
~:=
\begin{Diagram}{0}{-0.1}
\Duobox[2]{B}{0,0} \inwhite{A}{1.5} \bbmetric[0.7]{g}{-3.7,0}
\end{Diagram}
\end{equation}
We see that the effect of the hopping metric is to turn white dots into black dots.  Using these definitions we can write
\begin{equation}\label{boxesanddots}
\begin{Diagram}{0}{-0.1}
\Duobox[2]{A}{0,0} \bbinsert[0.8]{g}{4,0} \duosymbol{a} \Duobox[2]{B}{8,0}
\link{A}{g}{1.5}{1}\link{g}{B}{1}{1.5}
\end{Diagram}
~=~
\begin{Diagram}{0}{-0.1}
\Duobox[2]{A}{0,0} \Duobox[2]{B}{5,0}
\linkwb{A}{B}{1.5}{1.5}\duosymbol{a}
\end{Diagram}
~=~
\begin{Diagram}{0}{-0.1}
\Duobox[2]{A}{0,0} \Duobox[2]{B}{5,0}
\linkbw{A}{B}{1.5}{1.5}\duosymbol{a}
\end{Diagram}
~:=~
\begin{Diagram}{0}{-0.1}
\Duobox[2]{A}{0,0} \Duobox[2]{B}{4,0}
\link{A}{B}{1.5}{1.5}\duosymbol{a}
\end{Diagram}
\end{equation}
where last step is by definition.  Hence, we have
\begin{equation}
\begin{Diagram}{0}{-0.6}
\Opbox[2]{A}{0,0}\Opbox[2]{B}{0,4} \wire{A}{B}{1.5}{1.5}\opsymbol{a}
\end{Diagram}
~ \equiv ~
\begin{Diagram}{0}{0}
\Duobox[2]{A}{0,0} \Duobox[2]{B}{4,0}
\link{A}{B}{1.5}{1.5}\duosymbol{a}
\end{Diagram}
\end{equation}
and therefore (using (\ref{AequivtoprobA}))
\begin{equation}
\text{Prob}\left(\,
\begin{Diagram}{0}{-0.6}
\Opbox[2]{A}{0,0}\Opbox[2]{B}{0,4} \wire{A}{B}{1.5}{1.5}\opsymbol{a}
\end{Diagram}
\,\,\right)
~ = ~
\begin{Diagram}{0}{0}
\Duobox[2]{A}{0,0} \Duobox[2]{B}{4,0}
\link{A}{B}{1.5}{1.5}\duosymbol{a}
\end{Diagram}
\end{equation}
We see that the probability is given by a diagram that looks the same as the diagram providing the operational description of the experiment (except for the fact that we have used different font and directions for disambiguation).  We will see that, given Assumption 2 below, this is true in general.

\subsection{Black and white dots}

Examination of (\ref{boxesanddots}) reveals that an adjacent black and white dot can cancel one another out.  It is possible to define quantities such that we can insert pairs of black and white dots on links and cancel them out as we wish while maintaining equality between the expressions.   In accordance with the rule that we can cancel black and white dot pairs we must have
\begin{equation}
\nwdots\bwdots = \nwdots   ~~~~~ \bwdots\bndots = \bndots ~~~~~ \nbdots\wbdots = \nbdots ~~~~~ \wbdots\wndots = \wndots
\end{equation}
Hence both $\bwdots$ and $\wbdots$ behave as the identity and so we can achieve consistency if we define them to be equal to the identity.
We also have
\begin{equation}
\bbdots\wwdots = \bwdots  ~~~~~~ \wwdots\bbdots = \wbdots
\end{equation}
As $\bwdots$ and $\wbdots$ are both equal to the identity, we can achieve consistency if we define $\wwdots$ to be equal to the inverse of $\bbdots$.  With these definitions we can create  or cancel as many black/white or white/black pairs of dots along a link as we wish.  So long as the black and white dots are matched along the link (no same colour pairs) then we can always simplify down to a link without any dots.  This also means we can swap the order of the dots:
\begin{equation}
\nwdots\bndots = \nndotslong = \nbdots\wndots
\end{equation}
This can be useful in diagrammatic calculations.

One interesting consistency check is the following.  We have
\begin{equation}\label{whitedotfids}
\begin{Diagram}{0}{0.1}
\fideffect{X}{0,0} \outblack{X}{1}\Duosymbol[65,0]{a} \inwire{X}{1}\Opsymbol{a}
\wwmetric[0.7]{g}{3.1,7pt}
\end{Diagram}
~=~
\begin{Diagram}{0}{0.1}
\fideffect{X}{0,0} \outwhite{X}{1}\Duosymbol{a} \inwire{X}{1}\Opsymbol{a}
\end{Diagram}
~~~~~~~~~~
\begin{Diagram}{0}{-0.1}
\fidprep{X}{0,0} \inblack{X}{1} \Duosymbol[-65,0]{a} \outwire{X}{1} \opsymbol{a}
\wwmetric[0.7]{g}{-3.1,-7pt}
\end{Diagram}
~=~
\begin{Diagram}{0}{-0.1}
\fidprep{X}{0,0} \inwhite{X}{1} \Duosymbol{a} \outwire{X}{1} \opsymbol{a}
\end{Diagram}
\end{equation}
Hence, using (\ref{defbbequiv}),
\begin{equation}\label{wwhoppingmetric}
\begin{Diagram}{0}{-0.6}
\fidprep{X2}{4,0}
\fideffect{X}{4,4}
\wire{X2}{X}{1}{1}\opsymbol{a}
\inwhite{X2}{1} \Duosymbol{a}
\outwhite{X}{1}\Duosymbol{a}
\end{Diagram}
~\equiv ~
a \wwdots\bbdots\wwdots a =  a \wwdots a
\end{equation}
as we expect.

\subsection{Symbolic calculation for simple circuit}

We will now repeat the calculation for this simple circuit in symbolic notation.  Thus we have
\begin{equation}\label{probAB}
\mathsf{A^{a_1}B_{a_1}}  \equiv  {}_{a'_1}\!g^{a_1}\,\,{}^{a'_1}\!\!A B_{a_1} = A^{a_1}B_{a_1}= {}^{a_1}\!\!A \,\,{}_{a_1}\!B
\end{equation}
(c.f.\ (\ref{ABdiagramcalculation})).  Here we define
\begin{equation}
{}_{a'_1}g^{a_1} := p(\,{}_{a'_1}\!\mathsf{X}^\mathsf{a_1} \mathsf{X}_\mathsf{a_1}^{a_1})
\end{equation}
This is  the hoppping metric, \bbdots, in symbolic notation (see (\ref{defbb})).  We also define
\begin{equation}
A^{a_1} :=  {}_{a'_1}\!g^{a_1}\,\,{}^{a'_1}\!\!A  ~~~~~~~  {}_{a_1}\!B := {}_{a_1}\!g^{a'_1}\,\, B_{a'_1}
\end{equation}
(compare with (\ref{ABhoppingdiagram})).
Hence, $A_{a_1}$ and $B_{a_1}$ behave as dual vectors and must have the same number of components (this proves that we have the same number of fiducial preparations and effects for a given type).
We see that ${}_{a'_1}g^{a_1}$ causes a superscript to hop from right to left and a subscript to jump from left to right.  We define ${}^{a'_1}g_{a_1}$ so that it causes the indices to hop in the opposite direction.  Hence
\begin{equation}\label{hoppingeqns}
{}_{a''_1}g^{a_1} {}^{a''_1}g_{a'_1} := g^{a_1}_{a'_1} := \delta^{a_1}_{a'_1} ~~~~ \text{and} ~~~~
{}_{a_1}g^{a''_1} {}^{a'_1}g_{a''_1} := {}_{a_1}^{a'_1}g := {}_{a_1}^{a'_1}\delta
\end{equation}
We must have the delta function here since if we hop in one direction then in the other direction we must end up with the same expression we started with.  Therefore the entries of ${}^{a'_1}g_{a_1}$ are given by matrix inversion
\begin{equation}
{}^{a'_1}g_{a_1}=\big({}_{a'_1}g^{a_1}\big)^{-1}
\end{equation}
The correspondences between the symbolic and diagrammatic representations are
\begin{equation}
{}_{a'_1}g^{a_1}\Leftrightarrow \bbdots ~~~~
{}^{a'_1}g_{a_1}\Leftrightarrow \wwdots ~~~~
g^{a_1}_{a'_1}  \Leftrightarrow \wbdots ~~~~
{}_{a_1}^{a'_1}g \Leftrightarrow \bwdots
\end{equation}
We have introduced all these rules for the very simple circuit shown in (\ref{ABcircuit}).  We will see that, with assumption 2 to be introduced in the next section, these rules will apply to general circuits.

\section{Assumption 2}\label{assumption2section}

We now introduce our second assumption:
\begin{quote} {\bf Assumption 2: Operations are fully decomposable}.  We assume that any operation $\mathsf{A_{a_1b_2\dots c_3}^{d_4e_5\dots f_6}}$ can be written, in diagrammatic notation, as
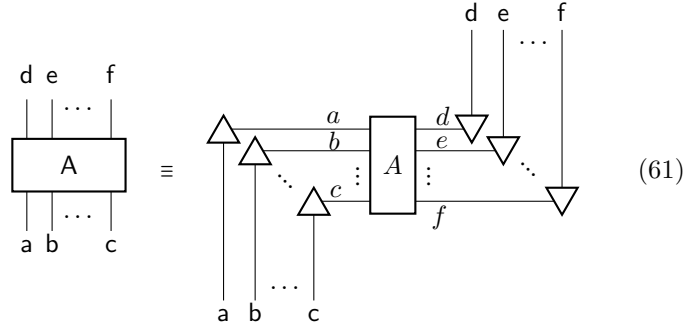
\begin{equation}\label{assumption2diagram}
\begin{Diagram}[1.4]{0}{0}
\Opbox[5]{A}{0,0}
\inwire{A}{1}\Opsymbol{a} \inwire{A}{2.2}\Opsymbol{b} \putlatex[30,15]{\ensuremath{\dots}} \inwire{A}{5}\Opsymbol{c}
\outwire{A}{1}\Opsymbol{d}\outwire{A}{2.2}\Opsymbol{e}\putlatex[30,-50]{\ensuremath{\dots}}   \outwire{A}{5}\Opsymbol{f}
\end{Diagram}
~~~\equiv ~~~
\begin{Diagram}[1.2]{0}{0}
\Duobox[5]{A}{0,0}
\putlatex[-44,-14]{\ensuremath{\vdots}} \putlatex[-135,-20]{\ensuremath{\ddots}}
\putlatex[45,-14]{\ensuremath{\vdots}} \putlatex[170,-5]{\ensuremath{\ddots}}
\linkedeffect[0.8]{A}{5}{c}{-3.5}{0}\duosymbol[-13,-4]{c} \thispoint{cbase}{-3.5,-6}\Opsymbol{c} \wire{cbase}{c}{1}{1}
\linkedeffect[0.8]{A}{2.2}{b}{-6.1}{0}\duosymbol[22,-3]{b}  \thispoint{bbase}{-6.1,-6}\Opsymbol{b} \wire{bbase}{b}{1}{1}
\linkedeffect[0.8]{A}{1}{a}{-7.5}{0}\duosymbol[41,-1]{a}  \thispoint{abase}{-7.5,-6}\Opsymbol{a} \wire{abase}{a}{1}{1}
\placelatex[-20, 16]{-4,-6}{\ensuremath{\dots}}
\linkedprep[0.8]{A}{1}{d}{3.5}{0}\duosymbol[4,1]{d}  \thispoint{dbase}{3.5,6}\Opsymbol[0,38]{d} \wire{d}{dbase}{1}{1}
\linkedprep[0.8]{A}{2.2}{e}{4.9}{0}\duosymbol[-17,-5]{e} \thispoint{ebase}{4.9,6}\Opsymbol[0,38]{e} \wire{e}{ebase}{1}{1}
\linkedprep[0.8]{A}{5}{f}{7.5}{0}\otherside\duosymbol[-57,-4]{f}  \thispoint{fbase}{7.5,6}\Opsymbol[0,38]{f} \wire{f}{fbase}{1}{1}
\placelatex[9, -16]{6,6}{\ensuremath{\dots}}
\end{Diagram}
\end{equation}
or, in symbolic notation,
\begin{equation}\label{keyassumption}
\mathsf{A_{a_1b_2\dots c_3}^{d_4e_5\dots f_6}} \,\,\equiv {}^{d_4e_5\dots f_6}\!A_{a_1b_2\dots c_3}\,\, \mathsf{X}_\mathsf{a_1}^{a_1} \mathsf{X}_\mathsf{b_2}^{b_2} \cdots \mathsf{X}_\mathsf{c_3}^{c_3} \,\,{}_{d_4}\!\mathsf{X}^\mathsf{d_4}{}_{e_5}\!\mathsf{X}^\mathsf{e_5}\cdots {}_{f_6}\!\mathsf{X}^\mathsf{f_6}
\end{equation}
In words we will say that any operation is equivalent to a linear combination of operations each of which consists of an effect for each input and a preparation for each output.  We do not lose any generality by choosing these to be fiducial sets (as in (\ref{keyassumption},
\ref{assumption2diagram})) since any other set could be written as a linear combination of the fiducial set.
\end{quote}
We allow the possibility that the entries in ${}^{d_4e_5\dots f_6}\!A_{a_1b_2\dots c_3}$ are negative (and this will, indeed, be the case in quantum theory).  Hence, in general, this cannot be thought of as physical mixing.  If we insert black and white dots in the links in (\ref{assumption2diagram}) such that we have black dots on the fiducial effects and preparations (as they were originally defined) then we get the duotensor
\begin{equation}
\begin{Diagram}{0}{0}
\Duobox[5]{A}{0,0}
\putlatex[-44,-14]{\ensuremath{\vdots}} \putlatex[44,-14]{\ensuremath{\vdots}}
\inwhite{A}{1}\Duosymbol{a}\inwhite{A}{2.2}\Duosymbol{b} \inwhite{A}{5}\Duosymbol{c}
\outwhite{A}{1}\Duosymbol{d}\outwhite{A}{2.2}\Duosymbol{e} \outwhite{A}{5}\Duosymbol{f}
\end{Diagram}
\end{equation}
with all white dots.

Assumption 2 introduces a subtly different attitude than the usual one concerning how we think about what an operation is. Usually we think of operations as effecting a transformation on systems as they pass through.  Here we think of an operation as corresponding to a bunch of separate effects and preparations. We need not think of systems as things that preserve their identity as they pass through - we do not use the same labels for wires coming out as going in.  This is certainly a more natural attitude when there can be different numbers of input and output systems and when they can be of different types.  Both classical and quantum transformations satisfy this assumption.  In spite of the different attitude just mentioned, we can implement arbitrary transformations, such as unitary transformation in quantum theory, by taking an appropriate sum over such effect and preparation operations.

\section{Duotensors}

We can place black and white dots on the links in (\ref{assumption2diagram}) in accordance with the meanings given to them in Sec.\ \ref{simplecircuit}.  In this way we can extract a box with inputs and outputs having black and white dots.  For example
\begin{equation}
\begin{Diagram}{0}{0}
\Duobox[4]{A}{0,0}
\outwhite[-4]{A}{1.5}\Duosymbol{a} \outwhite{A}{2.5}\Duosymbol{b} \outblack[4]{A}{3.5}\Duosymbol{d}
\inblack[-6]{A}{1}\Duosymbol{b}    \inblack[-2]{A}{2}\Duosymbol{c}   \inwhite[2]{A}{3}\Duosymbol{b} \inwhite[6]{A}{4}\Duosymbol{c}
\end{Diagram}
\end{equation}
In symbolic notation this corresponds to
\begin{equation}
  {}^{a_1b_2}_{b_3c_4} \! A^{d_5}_{b_6c_7}
\end{equation}
This object is tensor-like with a bit more structure, indices can appear on the left as well as the right.  The reason for this is that there are two independently chosen basis sets associated with every index - a fiducial set of effects and a fiducial set of preparations.  (For tensors we only have one choice of basis set associated with each index.)  Given this, we will call this mathematical object a {\it duotensor}.  Correspondingly, we can put an index on the right or hop it over to the left (using the hopping tensor), or vice versa.  For example,
\begin{equation}
\begin{Diagram}{0}{-0.1}
\Duobox[4]{A}{0,0}
\outwhite[-4]{A}{1.5}\Duosymbol{a} \outblack{A}{2.5}\Duosymbol{b} \outblack[4]{A}{3.5}\Duosymbol{d}
\inblack[-6]{A}{1}\Duosymbol{b}    \inwhite[-2]{A}{2}\Duosymbol{c}   \inblack[2]{A}{3}\Duosymbol{b} \inwhite[6]{A}{4}\Duosymbol{c}
\end{Diagram}
\!\! ~=~ \,
\begin{Diagram}{0}{-0.1}
\Duobox[4]{A}{0,0}
\outwhite[-4]{A}{1.5}\Duosymbol{a} \outwhite{A}{2.5}\Duosymbol[70,0]{b} \outblack[4]{A}{3.5}\Duosymbol{d}
\inblack[-6]{A}{1}\Duosymbol{b}    \inblack[-2]{A}{2}\Duosymbol[-70,0]{c}   \inwhite[2]{A}{3}\Duosymbol[-70,0]{b} \inwhite[6]{A}{4}\Duosymbol{c}
\wwmetric[0.8]{g1}{-3.8,0.53} \bbmetric[0.8]{g2}{-3.8,-0.53} \bbmetric[0.8]{g3}{3.8,0}
\end{Diagram}
\end{equation}
or, in symbolic form,
\begin{equation}
{}^{a_1b_2}_{b_3b_6} \! A^{d_5}_{c_4c_7}   = {}_{b'_6} g^{b_2}\,{}_{b_6} g^{b'_6}\, {}^{c'_4}\! g_{c_4}\,\, {}^{a_1b_2}_{b_3c'_4} \! A^{d_5}_{b'_6c_7}
\end{equation}
Subscripts always correspond to inputs and superscripts always correspond to outputs.  For the diagrammatic representation, subscripts go on left and superscripts on the right of the boxes.  The map between black and white dots and the placement of indices is given by
\begin{equation}
{}_\bullet^\circ A^\bullet_\circ
\end{equation}
We must match black and white dots.  Correspondingly, when we have a repeated index, we can either sum over a subscript and superscript, or over a pre-superscript and pre-subscript.  We cannot sum over indices that are on opposite sides. In the case that we want to sum over duotensors that are not matched like this we can always use the hopping metric or its inverse to make the indices match.

For an object to be a duotensor it must transform appropriately under transformation of the fiducial preparations and effects. We will indicate the original preparations and effects by $\mathsf X$ and the new set by $\tilde{\mathsf X}$.  Then we can write the old in terms of the new. For effects we have
\begin{equation}
\mathsf{X}_\mathsf{a_1}^{a_1} =   \mathcal{E}^{a_1}_{\tilde{a}_1} \,\, \mathsf{\tilde{X}}_\mathsf{a_1}^{\tilde{a}_1}
\end{equation}
where $\mathcal{E}^{a_1}_{\tilde{a}_1}$ is the transformation matrix for effects.  For preparations we have
\begin{equation}
{}_{a_1}\!\mathsf{X}^\mathsf{a_1} =   {}^{\tilde{a}_1}_{a_1}\! \mathcal{P}  \,\, {}_{\tilde{a}_1}\!\mathsf{\tilde{X}}^\mathsf{a_1}
\end{equation}
where ${}_{\tilde{a}_1}^{a_1}\!\mathcal{P}$ is the transformation matrix for preparations.  Consider
\begin{eqnarray}
\mathsf{A_{a_1b_2}^{c_3d_4}} \,\,& = & {}^{c_3d_4}\!A_{a_1b_2}\,\, \mathsf{X}_\mathsf{a_1}^{a_1} \mathsf{X}_\mathsf{b_2}^{b_2}  \,\,{}_{c_3}\!\mathsf{X}^\mathsf{c_3}{}_{d_4}\!\mathsf{X}^\mathsf{d_4}   \nonumber\\
& {} & \nonumber \\
& = & {}^{\tilde{c}_3\tilde{d}_4}\!\tilde{A}_{\tilde{a}_1\tilde{b}_2}\,\, \mathsf{\tilde{X}}_\mathsf{{a}_1}^{\tilde{a}_1} \mathsf{\tilde{X}}_\mathsf{b_2}^{\tilde{b}_2}  \,\,{}_{\tilde{c}_3}\!\mathsf{\tilde{X}}^\mathsf{c_3}{}_{\tilde{d}_4}\!\mathsf{\tilde{X}}^\mathsf{d_4}
\nonumber
\end{eqnarray}
Clearly
\begin{equation}\label{presupsubtrans}
{}^{\tilde{c}_3\tilde{d}_4}\!\tilde{A}_{\tilde{a}_1\tilde{b}_2} =      {}^{c_3d_4}\!A_{a_1b_2} \,\,  \mathcal{E}^{a_1}_{\tilde{a_1}}\,\, \mathcal{E}^{b_2}_{\tilde{b_2}}   \,\, {}_{c_3}^{\tilde{c}_3} \mathcal{P} \,\,{}_{d_4}^{\tilde{d}_4} \mathcal{P}
\end{equation}
This equation shows how a duotensor transforms if it has only pre-superscripts and subscripts.  To see how it transforms if we have indices in other positions we note, using (\ref{probAB}) and transforming the subscript as in (\ref{presupsubtrans})
\begin{equation}
\text{Prob}(\mathsf{A^{a_1} B_{a_1}}) = A^{a_1} B_{a_1} = \tilde{A}^{\tilde{a}_1} \tilde{B}_{\tilde{b}_1} = \tilde{A}^{\tilde{a}_1}\,\, \mathcal{E}^{a_1}_{\tilde{a_1}}\, B_{a_1}
\end{equation}
This equation must hold for any $B^{a_1}$.  Hence we must have
\begin{equation}
 A^{a_1} = \tilde{A}^{\tilde{a}_1}\,\, \mathcal{E}^{a_1}_{\tilde{a_1}}
\end{equation}
or
\begin{equation}
\tilde{A}^{\tilde{a}_1} =  A^{a_1} \,\, \mathcal{E}_{a_1}^{\tilde{a_1}}
\end{equation}
where $\mathcal{E}_{a_1}^{\tilde{a_1}}$ is the inverse of $\mathcal{E}^{a_1}_{\tilde{a_1}}$ such that
\begin{equation}
\mathcal{E}^{a_1}_{\tilde{a_1}} \, \mathcal{E}_{a_1}^{\tilde{a'_1}} = \delta_{\tilde{a}_1}^{\tilde{a}'_1}
\end{equation}
Hence superscripts on duotensors transform with $\mathcal{E}_{a_1}^{\tilde{a_1}}$.

By considering
\begin{equation}
\text{Prob}(\mathsf{A^{a_1} B_{a_1}}) = {}^{a_1}\!A \,\,\, {}_{a_1}\!B = {}^{\tilde{a}_1}\!\!\tilde{A}\,\,\, {}_{\tilde{b}_1}\!\!\tilde{B}
\end{equation}
and employing similar reasoning to that above, we can easily prove that pre-subscripts transform with  ${}^{a_1}_{\tilde{a}_1} \mathcal{P}$, this being is the inverse of ${}_{a_1}^{\tilde{a}_1} \mathcal{P}$, i.e.\
\begin{equation}
{}^{a_1}_{\tilde{a}'_1} \mathcal{P}\,\, {}_{a_1}^{\tilde{a}_1}\! \mathcal{P} = {}^{\tilde{a}_1}_{\tilde{a}'} \delta
\end{equation}
Hence, the transformation rule for a duotensor with indices in all positions is illustrated by
\begin{equation}\label{duotensortrans}
{}^{\tilde{c}_3}_{\tilde{d}_4}\!\tilde{A}_{\tilde{a}_1}^{\tilde{b}_2} =      {}^{c_3}_{d_4}\!A_{a_1}^{b_2} \,\,\,\,  \mathcal{E}^{a_1}_{\tilde{a_1}}\,\, \, \mathcal{E}_{b_2}^{\tilde{b_2}}   \,\,\, {}_{c_3}^{\tilde{c}_3} \mathcal{P} \,\,\,{}^{d_4}_{\tilde{d}_4} \mathcal{P}
\end{equation}
We see that a duotensor that has only subscripts and superscripts transforms as a tensor with respect to the transformation matrix for effects.  A duotensor that has only pre-superscripts and pre-subscripts transforms as a tensor with respect to the transformation matrix for preparations.  However, a duotensor with indices in all positions behaves like a new object that transforms with transformation matrices for the effects and the preparations.  Further, there exists a hopping metric which can take indices from the left to the right and vice-versa.  The duotensor is a generalization of the idea of a tensor.  It has particular application to operational probabilistic theories. We should note that we have a choice of fiducial effects and fiducial preparations for each type.  In general we do not expect $K_{\mathsf a}$ and $K_{\mathsf b}$ to be equal.  Hence the indices for different types will, in general, run over different numbers of values.  This is different from the tensors used in General Relativity where all indices run over the four dimensions of space-time.

There is a certain notational difficulty associated with the symbolic representation of duotensors (though not the diagrammatic representation).  This is that, when we hop indices over, it is not clear what position they should occupy on the other side.  This could be solved with more heavy handed notation. For example, we could attempt to space the indices in accordance with their positions leaving gaps where there is no index on that side.  Alternatively,  we could include a superscript on each index (in addition to the subscript) to indicate which position it goes in.  For example,  $a_5^2$ would go in position 2.  This notational difficulty is avoided if we keep the duotensor in tensor form with all the  indices on the right hand side (we could have chosen the left hand side). We will call this the {\it standard form}.   We can sum over duotensors in standard form without using the hopping metric or its inverse.  Hence, this form allows us to calculate general expressions.  However, if duotensors were only ever presented in standard form we would lose some of the natural structure of the mathematics. It can be of interest to use a non-standard form. For example, in Sec.\ \ref{sigofduotensors} we will see that the form of the duotensor having all black dots (presubscripts and superscripts) is equal to the probabilities obtained if have fiducial preparations on all the inputs and fiducial effects on all the outputs.

\section{General circuits}

We will show how to calculate the probability for a general circuit by means of an example.   We will work in diagrammatic notation.  The same calculation could  be done in symbolic notation.  Consider the circuit
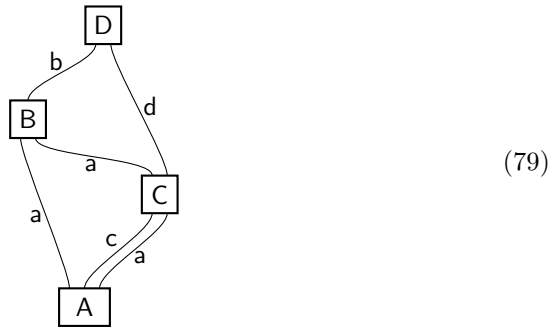
\begin{equation}
\begin{Diagram}{0}{-2}
\Opbox{A}{0,0} \Opbox[2]{C}{4,6} \Opbox[2]{B}{-3,10} \Opbox[2]{D}{1,15}
\wire{A}{B}{1}{1}\opsymbol{a} \wire{A}{C}{2}{1}\opsymbol[0,6]{c} \wire{A}{C}{3}{2}\otherside\opsymbol{a} \wire{C}{B}{1}{2}\opsymbol{a}
\wire{C}{D}{2}{2}\otherside\opsymbol[4,0]{d} \wire{B}{D}{1.5}{1}\opsymbol[0,6]{b}
\end{Diagram}
\end{equation}
Using Assumption 2 we see this is equivalent to
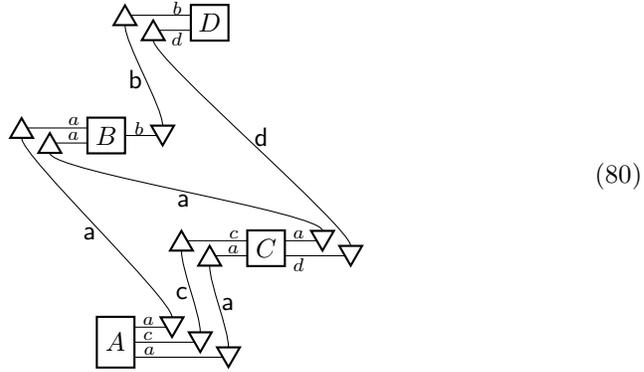
\begin{equation}
\begin{Diagram}{0}{-3.3}
\begin{move}{-2,4}
\Duobox{A}{0,0}
\linkedprep[0.7]{A}{1}{A1}{3}{0}\duosymbol[-3,-6]{\scriptstyle a}
\linkedprep[0.7]{A}{2}{A2}{4.5}{0}\duosymbol[-25,-6]{\scriptstyle c}
\linkedprep[0.7]{A}{3}{A3}{6}{0}\duosymbol[-44,-6]{\scriptstyle a}
\end{move}
\begin{move}{-2.5,15}
\Duobox[2]{B}{0,0}
\linkedeffect[0.7]{B}{1}{1B}{-4.5}{0}\duosymbol[25,-6]{\scriptstyle a}
\linkedeffect[0.7]{B}{2}{2B}{-3}{0}\duosymbol[3,-6]{\scriptstyle a}
\linkedprep[0.7]{B}{1.5}{B15}{3}{0} \duosymbol[-3,-6]{\scriptstyle b}
\end{move}
\begin{move}{6,9}
\Duobox[2]{C}{0,0}
\linkedeffect[0.7]{C}{1}{1C}{-4.5}{0}\duosymbol[25,-6]{\scriptstyle c}
\linkedeffect[0.7]{C}{2}{2C}{-3}{0}\duosymbol[3,-6]{\scriptstyle a}
\linkedprep[0.7]{C}{1}{C1}{3}{0} \duosymbol[-3,-6]{\scriptstyle a}
\linkedprep[0.7]{C}{2}{C2}{4.5}{0} \otherside\duosymbol[-25,3]{\scriptstyle d}
\end{move}
\begin{move}{3,21}
\Duobox[2]{D}{0,0}
\linkedeffect[0.7]{D}{1}{1D}{-4.5}{0}\duosymbol[25,-6]{\scriptstyle b}
\linkedeffect[0.7]{D}{2}{2D}{-3}{0}\otherside\duosymbol[3,3]{\scriptstyle d}
\end{move}
\wire{A1}{1B}{1}{1}\opsymbol{a} \wire{A2}{1C}{1}{1}\opsymbol{c} \wire{A3}{2C}{1}{1}\otherside\opsymbol{a}
\wire{C1}{2B}{1}{1}\opsymbol{a} \wire{C2}{2D}{1}{1}\otherside\opsymbol[4,0]{d} \wire{B15}{1D}{1}{1}\opsymbol{b}
\end{Diagram}
\end{equation}
This diagram is a linear sum of circuits.  Each circuit in this linear sum consists of six (in this case) disjoint circuits each made out of a fiducial preparation followed by a fiducial effect (we will call these {\it fiducial circuits}). To make the next step clearer we will distort this graph (which does not change its meaning)
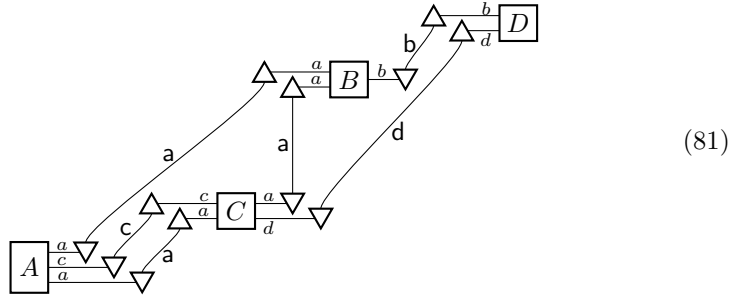
\begin{equation}
\begin{Diagram}{0}{-3}
\begin{move}{-7,5}
\Duobox{A}{0,0}
\linkedprep[0.7]{A}{1}{A1}{3}{0}\duosymbol[-3,-6]{\scriptstyle a}
\linkedprep[0.7]{A}{2}{A2}{4.5}{0}\duosymbol[-25,-6]{\scriptstyle c}
\linkedprep[0.7]{A}{3}{A3}{6}{0}\duosymbol[-44,-6]{\scriptstyle a}
\end{move}
\begin{move}{10,15}
\Duobox[2]{B}{0,0}
\linkedeffect[0.7]{B}{1}{1B}{-4.5}{0}\duosymbol[25,-6]{\scriptstyle a}
\linkedeffect[0.7]{B}{2}{2B}{-3}{0}\duosymbol[3,-6]{\scriptstyle a}
\linkedprep[0.7]{B}{1.5}{B15}{3}{0} \duosymbol[-3,-3]{\scriptstyle b}
\end{move}
\begin{move}{4,8}
\Duobox[2]{C}{0,0}
\linkedeffect[0.7]{C}{1}{1C}{-4.5}{0}\duosymbol[25,-6]{\scriptstyle c}
\linkedeffect[0.7]{C}{2}{2C}{-3}{0}\duosymbol[3,-6]{\scriptstyle a}
\linkedprep[0.7]{C}{1}{C1}{3}{0} \duosymbol[-3,-6]{\scriptstyle a}
\linkedprep[0.7]{C}{2}{C2}{4.5}{0} \otherside\duosymbol[-25,3]{\scriptstyle d}
\end{move}
\begin{move}{19,18}
\Duobox[2]{D}{0,0}
\linkedeffect[0.7]{D}{1}{1D}{-4.5}{0}\duosymbol[25,-6]{\scriptstyle b}
\linkedeffect[0.7]{D}{2}{2D}{-3}{0}\otherside\duosymbol[3,3]{\scriptstyle d}
\end{move}
\wire{A1}{1B}{1}{1}\opsymbol{a} \wire[0.4]{A2}{1C}{1}{1}\opsymbol{c} \wire[0.4]{A3}{2C}{1}{1}\otherside\opsymbol{a}
\wire{C1}{2B}{1}{1}\opsymbol{a} \wire{C2}{2D}{1}{1}\otherside\opsymbol{d} \wire[0.4]{B15}{1D}{1}{1}\opsymbol[-3,0]{b}
\end{Diagram}
\end{equation}
If we insert black and white dots appropriately on each link (so the fiducial elements have a black dot) then use (\ref{defbbequiv}) we see that this is equivalent to
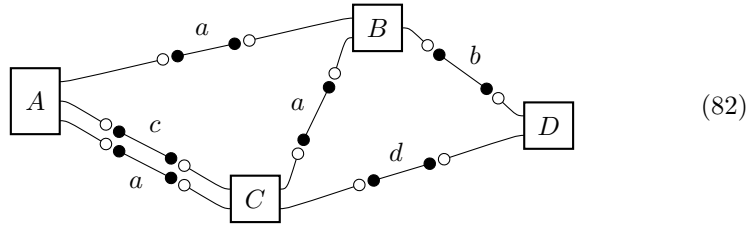
\begin{equation}
\begin{Diagram}[1.3]{0}{0}
\Duobox{A}{-3,0} \Duobox[2]{B}{11,3} \Duobox[2]{C}{6,-4}  \Duobox[2]{D}{18,-1}
\linkwbbw[0.6]{A}{B}{1}{1}\duosymbol{a} \linkwbbw[0.6]{A}{C}{2}{1}\duosymbol{c} \linkwbbw[0.6]{A}{C}{3}{2}\otherside\duosymbol{a}
\linkwbbw[0.6]{C}{B}{1}{2}\duosymbol{a} \linkwbbw[0.6]{C}{D}{2}{2}\duosymbol{d} \linkwbbw[0.6]{B}{D}{1.5}{1}\duosymbol{b}
\end{Diagram}
\end{equation}
Note that we are implicitly using the fact that the $p(\cdot)$ function factorizes over the six disjoint fiducial circuits in making this step.
Now canceling over each pair of black and white dots, we get
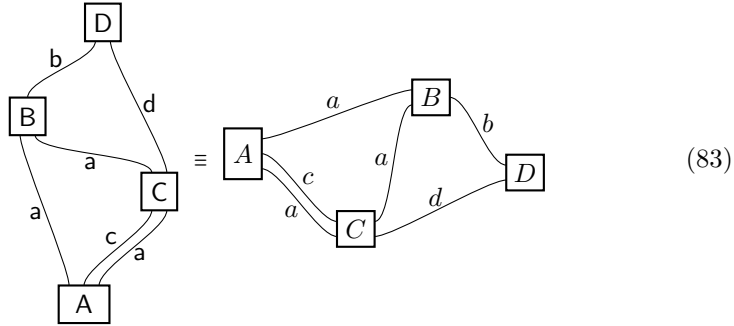
\begin{equation}
\begin{Diagram}{0}{-2}
\Opbox{A}{0,0} \Opbox[2]{C}{4,6} \Opbox[2]{B}{-3,10} \Opbox[2]{D}{1,15}
\wire{A}{B}{1}{1}\opsymbol{a} \wire{A}{C}{2}{1}\opsymbol[0,6]{c} \wire{A}{C}{3}{2}\otherside\opsymbol{a} \wire{C}{B}{1}{2}\opsymbol{a}
\wire{C}{D}{2}{2}\otherside\opsymbol[4,0]{d} \wire{B}{D}{1.5}{1}\opsymbol[0,6]{b}
\end{Diagram}
~ \equiv ~
\begin{Diagram}{0}{0}
\Duobox{A}{0,0} \Duobox[2]{B}{10,3} \Duobox[2]{C}{6,-4}  \Duobox[2]{D}{15,-1}
\link{A}{B}{1}{1}\duosymbol{a} \link{A}{C}{2}{1}\duosymbol{c} \link{A}{C}{3}{2}\otherside\duosymbol{a}
\link{C}{B}{1}{2}\duosymbol{a} \link{C}{D}{2}{2}\duosymbol{d} \link{B}{D}{1.5}{1}\duosymbol{b}
\end{Diagram}
\end{equation}
Which implies, using (\ref{AequivtoprobA}), that
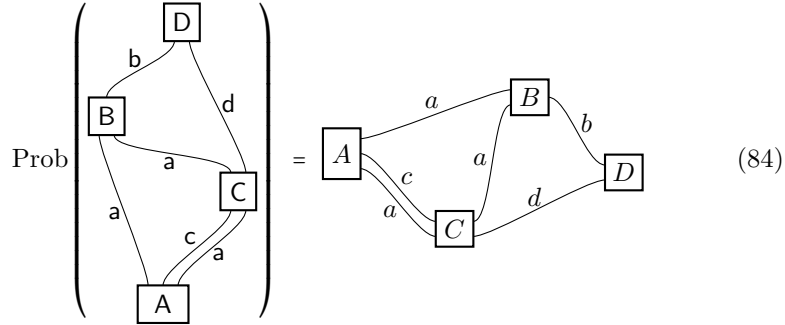
\begin{equation}
\text{Prob}\left(
\begin{Diagram}{0}{-2}
\Opbox{A}{0,0} \Opbox[2]{C}{4,6} \Opbox[2]{B}{-3,10} \Opbox[2]{D}{1,15}
\wire{A}{B}{1}{1}\opsymbol{a} \wire{A}{C}{2}{1}\opsymbol[0,6]{c} \wire{A}{C}{3}{2}\otherside\opsymbol{a} \wire{C}{B}{1}{2}\opsymbol{a}
\wire{C}{D}{2}{2}\otherside\opsymbol[4,0]{d} \wire{B}{D}{1.5}{1}\opsymbol[0,6]{b}
\end{Diagram}
\right)
~ = ~
\begin{Diagram}{0}{0}
\Duobox{A}{0,0} \Duobox[2]{B}{10,3} \Duobox[2]{C}{6,-4}  \Duobox[2]{D}{15,-1}
\link{A}{B}{1}{1}\duosymbol{a} \link{A}{C}{2}{1}\duosymbol{c} \link{A}{C}{3}{2}\otherside\duosymbol{a}
\link{C}{B}{1}{2}\duosymbol{a} \link{C}{D}{2}{2}\duosymbol{d} \link{B}{D}{1.5}{1}\duosymbol{b}
\end{Diagram}
\end{equation}
It is striking that
\begin{quote}
{\bf The diagram for the mathematical calculation looks the same as the diagram for the operational description}.
\end{quote}
Of course, the duotensor diagram is rotated by $90^\circ$ and the font is different but these are just conventions put in place to disambiguate the maths from the physics. It is clear from the above calculation that this will be true for any circuit.  We have here, then, a remarkable similarity between the structure of two very different things.  The operational description basically tells an experimentalist how to do the experiment, and the duotensor diagram tells a mathematician how to do the calculation.

In symbolic notation, we obtain a similar result.  The circuit we are considering is, in symbolic notation,
\begin{equation}
\mathsf{ A^{a_1c_2a_3} B_{a_1a_4}^{b_6} C_{c_2a_3}^{a_4d_5} D_{b_6d_5} }
\end{equation}
We can easilly show, by analogous reasoning to that for the diagrams, that
\begin{equation}
\mathsf{ A^{a_1c_2a_3} B_{a_1a_4}^{b_6} C_{c_2a_3}^{a_4d_5} D_{b_6d_5} }
\equiv  A^{a_1c_2a_3} B_{a_1a_4}^{b_6} C_{c_2a_3}^{a_4d_5} D_{b_6d_5}
\end{equation}
and consequently
\begin{equation}
\text{Prob}\left(\mathsf{ A^{a_1c_2a_3} B_{a_1a_4}^{b_6} C_{c_2a_3}^{a_4d_5} D_{b_6d_5} } \right)
=  A^{a_1c_2a_3} B_{a_1a_4}^{b_6} C_{c_2a_3}^{a_4d_5} D_{b_6d_5}
\end{equation}
Hence we see that the probability for a circuit is given by a duotensorial calculation that has the same symbolic representation as the operational description of the circuit has (except that the font is different).  Once again, this is clearly true for any circuit.

The translation between the physics (the operational description) and the mathematics (the duotensor calculation) is accomplished by application of Assumption 2 (though Assumption 1 also plays an important role).  Diagram (\ref{assumption2diagram}), for Assumption 2, is a hybrid between physics and maths.  It has objects in it which point to elements of the experimental world, and objects in it which point to elements of a mathematical calculation.  It seems reasonable to suppose that all physical theories will have hybrid expressions of this sort.  However, normally it is not clearly elucidated as to what this expression is.

\section{General fragments}

Consider the circuit
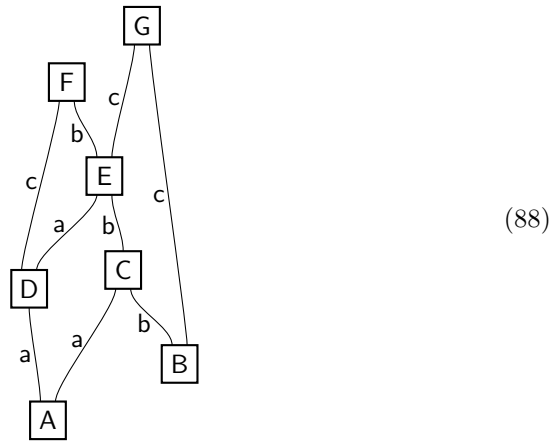
\begin{equation}\label{bigcircuit}
\begin{Diagram}{0}{-2}
\Opbox[2]{A}{3,-3}\Opbox[2]{D}{2,4}\Opbox[2]{F}{4,15}
\wire{A}{D}{1}{1.5}\opsymbol{a} \wire{D}{F}{1}{1}\opsymbol{c}
\Opbox[2]{C}{7,5}\Opbox[2]{E}{6,10}
\wire{C}{E}{1.5}{2}\opsymbol{b}
\Opbox[2]{B}{10,0}\Opbox[2]{G}{8,18}
\wire{B}{G}{2}{2}\opsymbol{c}
\wire{A}{C}{2}{1}\opsymbol{a} \wire{D}{E}{2}{1}\opsymbol{a} \wire{E}{F}{1}{2}\opsymbol{b}
\wire{B}{C}{1}{2}\opsymbol{b} \wire{E}{G}{2}{1}\opsymbol{c}
\end{Diagram}
\end{equation}
We can regard this as being made out of three fragments
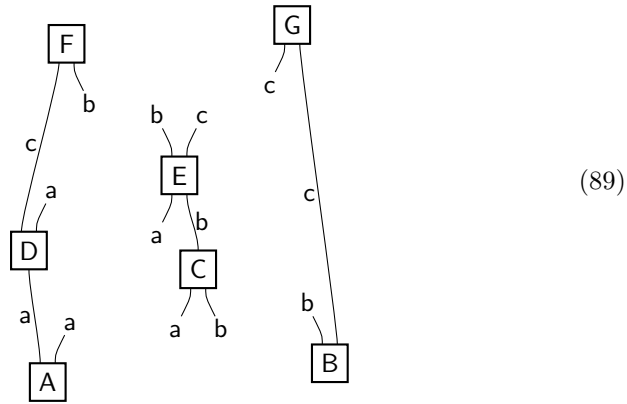
\begin{equation}
\begin{Diagram}{0}{-2}
\begin{move}{-4,0}
\Opbox[2]{A}{3,-3}\Opbox[2]{D}{2,4}\Opbox[2]{F}{4,15}
\wire{A}{D}{1}{1.5}\opsymbol{a} \wire{D}{F}{1}{1}\opsymbol{c}
\outwire[5]{A}{2} \Opsymbol{a} \outwire[5]{D}{2}\Opsymbol{a} \inwire[5]{F}{2}\Opsymbol{b}
\end{move}
\begin{move}{0,-2}
\Opbox[2]{C}{7,5}\Opbox[2]{E}{6,10}
\wire{C}{E}{1.5}{2}\otherside\opsymbol{b}
\inwire[-5]{C}{1}\Opsymbol{a} \inwire[5]{C}{2}\Opsymbol{b}
\inwire[-5]{E}{1}\Opsymbol{a} \outwire[-5]{E}{1}\Opsymbol{b} \outwire[5]{E}{2}\Opsymbol{c}
\end{move}
\begin{move}{4,-2}
\Opbox[2]{B}{10,0}\Opbox[2]{G}{8,18}
\wire{B}{G}{2}{2} \opsymbol{c}
\outwire[-5]{B}{1}\Opsymbol{b}
\inwire[-5]{G}{1}\Opsymbol{c}
\end{move}
%
%
\end{Diagram}
\end{equation}
This makes sense only if the outcome set for the whole circuit is the cartesian product of outcome sets for each of the fragments.  This is not much of an imposition since it is always true for the most fine-grained outcome sets.  These are the outcome sets that can be written as the cartesian product of one outcome for each operation.  We can certainly break circuits with such an outcome set up into fragments.  So one course of action is to work with such fine-grained outcome sets then course-grain later if we wish.  The course-graining process does not really add much in the way of physics so it would be fine to work all the way through with the most fine-grained outcome sets.
The circuit in (\ref{bigcircuit}) is equivalent to the duotensorial diagram
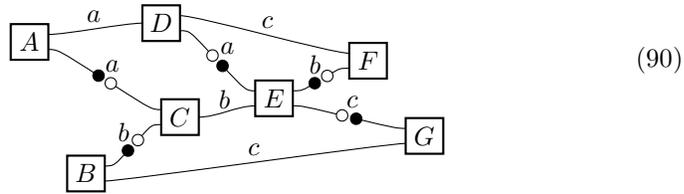
\begin{equation}\label{bigcircuitduotensor}
\begin{Diagram}{0}{0.9}
\Duobox[2]{A}{-3,-3}\Duobox[2]{D}{4,-2}\Duobox[2]{F}{15, -4}
\link{A}{D}{1}{1.5}\duosymbol{a} \link{D}{F}{1}{1}\duosymbol{c}
\Duobox[2]{C}{5,-7}\Duobox[2]{E}{10,-6}
\link{C}{E}{1.5}{2}\duosymbol{b}
\Duobox[2]{B}{0,-10}\Duobox[2]{G}{18,-8}
\link{B}{G}{2}{2} \duosymbol{c}
\linkbw{A}{C}{2}{1}\duosymbol{a} \linkwb{D}{E}{2}{1}\duosymbol{a} \linkbw{E}{F}{1}{2}\duosymbol{b}
\linkbw{B}{C}{1}{2}\duosymbol{b} \linkwb{E}{G}{2}{1}\duosymbol{c}
\end{Diagram}
\end{equation}
which therefore is equal to the probability for the circuit.  Note that we have arbitrarily inserted black and white dots on the links corresponding to the wires connecting the three fragments above.  We can break up this calculation into three separate calculations, one for each of the fragments
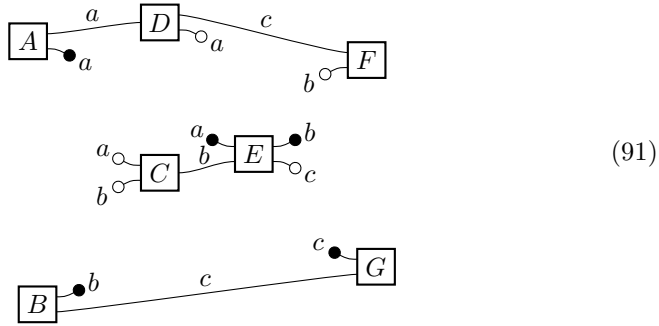
\begin{equation}
\begin{Diagram}{0}{1.4}
\begin{move}{1,3}
\Duobox[2]{A}{-3,-3}\Duobox[2]{D}{4,-2}\Duobox[2]{F}{15, -4}
\link{A}{D}{1}{1.5}\duosymbol{a} \link{D}{F}{1}{1}\duosymbol{c}
\outblack[5]{A}{2}\Duosymbol{a} \outwhite[5]{D}{2}\Duosymbol{a} \inwhite[5]{F}{2}\Duosymbol{b}
\end{move}
\begin{move}{0,0}
\Duobox[2]{C}{5,-7}\Duobox[2]{E}{10,-6}
\link{C}{E}{1.5}{2}\duosymbol{b}
\inwhite[-5]{C}{1}\Duosymbol{a}\inwhite[5]{C}{2}\Duosymbol{b}
\inblack[-5]{E}{1}\Duosymbol{a}
\outblack[-5]{E}{1}\Duosymbol{b} \outwhite[5]{E}{2}\Duosymbol{c}
\end{move}
\begin{move}{-1.5,-4}
\Duobox[2]{B}{0,-10}\Duobox[2]{G}{18,-8}
\link{B}{G}{2}{2} \duosymbol{c}
\outblack[-5]{B}{1}\Duosymbol{b} \inblack[-5]{G}{1}\Duosymbol{c}
\end{move}
\end{Diagram}
\end{equation}
Each of these objects is itself a duotensor (since they transform as duotensors).  Having calculated the duotensor associated with each fragment, we can calculate the probability for the whole circuit by summing over these fragment's duotensors as shown in (\ref{bigcircuitduotensor}).  Or we can just sum over two of them to get the duotensor for a bigger fragment.   The same fragment may appear in many different circuits of interest. Consequently it would be of use to calculate its duotensor and use it multiple times rather than having to calculate the whole circuit from its operations each time.

If we wish to course-grain our outcome sets for fragments, we can simply add duotensors corresponding to the fine-grained outcome sets together.  In this way, even if we had to fine-grain to apply the above techniques, we can get back to the level of course-graining we want to work at.

\section{The physical significance of the different forms of a duotensor}\label{sigofduotensors}

It can be shown that
\begin{equation}\label{allblackduotensor}
\begin{Diagram}{0}{0}
\Duobox[2]{A}{0,0}
\inblack{A}{1}\Duosymbol{a}\inblack{A}{2}\Duosymbol{b}
\outblack{A}{1}\Duosymbol{c}\outblack{A}{2}\Duosymbol{d}
\end{Diagram}
~=~
\text{Prob}\left(
\begin{Diagram}{0}{0}
\Opbox[2]{A}{0,0}
\fidprep[0.65]{1X}{-0.4,-3}\wire{1X}{A}{1}{1}\opsymbol{a}\inblack{1X}{1}\Duosymbol{a}
\fidprep[0.65]{2X}{0.4,-4.5}\wire{2X}{A}{1}{2}\otherside\opsymbol{b}\inblack{2X}{1}\Duosymbol{b}
\fideffect[0.65]{X1}{-0.4,4.5}\wire{A}{X1}{1}{1}\opsymbol{c}\outblack{X1}{1}\Duosymbol{c}
\fideffect[0.65]{X2}{0.4,3}\wire{A}{X2}{2}{1}\otherside\opsymbol{d}\outblack{X2}{1}\Duosymbol{d}
\end{Diagram}
\right)
\end{equation}
To see this first insert a black and white dot in each link on the LHS of the equation (with the white dot closer to box $\mathsf A$), then use (\ref{defbbequiv}), then (\ref{assumption2diagram}), and finally use (\ref{AequivtoprobA}).  Hence all the entries in this duotensor must be positive (between 0 and 1).  This works for fragments in general.  Hence if a duotensor corresponding to a fragment has all black dots then its entries are positive and equal to the fiducial probabilities (that is the probabilities obtained by putting fiducial preparations on inputs and fiducial effects on outputs).   We can put white dots on such a duotensor by introducing \wwdots.
\begin{equation}\label{puttingonwhitedots}
\begin{Diagram}{0}{0}
\Duobox[2]{A}{0,0}
\inwhite{A}{1}\Duosymbol{a}\inblack{A}{2}\Duosymbol{b}
\outblack{A}{1}\Duosymbol{c}\outwhite{A}{2}\Duosymbol{d}
\end{Diagram}
~=~
\begin{Diagram}{0}{0}
\Duobox[2]{A}{0,0}
\inblack{A}{1}\Duosymbol[-95,0]{a}\inblack{A}{2}\Duosymbol{b}
\outblack{A}{1}\Duosymbol{c}\outblack{A}{2}\Duosymbol[95,0]{d}
\wwmetric{1A}{-4.2,0.4}
\wwmetric{A2}{4.2,-0.4}
\end{Diagram}
\end{equation}
Recall \wwdots is the inverse of \bbdots.  We know that the entries of \bbdots are all probabilities from its definition (and since it has only black dots).  It is possible that its inverse will have negative entries.  This is, indeed, the case in quantum theory (though, in classical probability theory, \wwdots has positive entries).  Hence, the duotensor can become negative once it has white dots on it.  We might ask why not work only with duotensors having only black dots.  The answer is that we need the white dots to be able to join the duotensors up since we must match black and white dots.

The statement in (\ref{allblackduotensor}) is for operations.  Operations are examples of fragments.  In fact it is clear that a similar statement holds for general fragments. If we take the duotensor with all black dots corresponding to any fragment then this is equivalent to that fragment completed into a circuit with fiducial elements.

\section{Calculating well conditioned probability ratios}

We are now in a position to say, as was our stated objective in Sec. \ref{ourobjective}, when a probability ratio is well conditioned, and what it is equal to in the cases where it is well conditioned.  First we state the result.
\begin{quote} {\bf The probability ratio}
\begin{equation}\label{wellconditionedconditioncandidate}
\frac{\text{Prob}(\mathsf{E[i]})}{\text{Prob}(\mathsf{E[j]})}
\end{equation}
where $\mathsf{E[i]}$ and $\mathsf{E[j]}$ are two fragments corresponding to different outcome sets for the same experiment
is
\begin{description}
\item[well conditioned] if and only if the corresponding duotensors, $E[i]$ and $E[j]$, are proportional, and
\item[equal to] the constant of proportionality $k$ in $E[i]=k E[j]$ (if well conditioned).
\end{description}
\end{quote}
To prove this we note that, from Sec. \ref{probabilityratio}, for the probability ratio (\ref{wellconditionedconditioncandidate}) to be well conditioned, we require that
\begin{equation}\label{wellconditionedcondition}
\frac{\text{Prob}(\mathsf{E[i]F})}{\text{Prob}(\mathsf{E[j]F})}
\end{equation}
be independent of $\mathsf F$ for any choice of fragment $\mathsf F$ that completes the circuit.  One set of fragments for $\mathsf F$ we can consider are the fragments that consist of simply putting fiducial preparations on each input of $\mathsf{ E[i]}$, and $\mathsf{ E[j]}$, and fiducial effects on each output.  This gives us the fiducial probabilities for these two fragments.   Let us consider
(\ref{wellconditionedcondition}) with respect to these choices for $\mathsf F$.  We saw in Sec.\ \ref{sigofduotensors} that the entries of the duotensor with all black dots is equal to the fiducial probabilities.  Hence, for (\ref{wellconditionedcondition}) to hold, we require that the elements of $ E[i]$ are proportional to the corresponding elements of $ E[j]$ with the same constant of proportionality.  Hence the two duotensors must be proportional.  (This is clearly necessary when these two duotensors are in \lq\lq all black dots form" but it must also be true when they are both in any other given form since multiplication by hopping tensors, which are non-singular, will not effect such a proportionality relationship).   This is a necessary condition but it is also a sufficient condition since, clearly, (\ref{wellconditionedcondition}) is independent of the choice of $\mathsf F$ (where this completes the circuit) when $E[i]$ and $E[j]$ are proportional.   Further, when the two duotensors are parallel then the probability ratio is simply given by the proportionality constant.

This result is in accordance with our objectives as stated in Sec.\ \ref{ourobjective}.  In particular, note that we have the property of formalism locality since we employ only the duotensors associated with the fragments $E[i]$ and $E[j]$.

Given that $k$ is the constant of proportionality between $E[i]$ and $E[j]$ when they are proportional, it is tempting to write
\begin{equation}\label{probratioruleequiv}
\frac{\mathsf{E[i]}}{\mathsf{E[j]}} \equiv \frac{E[i]}{E[j]}
\end{equation}
In fact we can extend our notion of equivalence to ratios such that this expression is true even when $E[i]$ and $E[j]$ are not proportional.
In general we will say that
\begin{equation}
\frac{\text{expression}_1}{\text{expression}_2} \equiv \frac{\text{expression}_3}{\text{expression}_4}
\end{equation}
if
\begin{equation}
\frac{p(\text{expression}_1 \text{expression}_L)}{p(\text{expression}_2 \text{expression}_L)}
\equiv \frac{p(\text{expression}_3 \text{expression}_R)}{p(\text{expression}_4 \text{expression}_R)}
\end{equation}
for any $\text{expression}_{L,R}$.
Here, $\text{expression}_L$ is any expression that makes the contents of the arguments of the numerator and denominator on the left hand side into a linear sum of circuits.   On the right hand side $\text{expression}_R$ does the same thing.  Given this, and by the reasoning of Sec.\ \ref{equivalencerelations}, it is clear that (\ref{probratioruleequiv}) holds in general (not just when $E[i]$ and $E[j]$ are proportional).  Further, in the case that $E[i]$ and $E[j]$ are proportional, we can cancel down so that their ratio is equivalent to a simple ratio.  Hence we can restate the rule as follows.  The probability ratio (\ref{wellconditionedconditioncandidate}) is well defined if and only if
\begin{equation}
\frac{\mathsf{E[i]}}{\mathsf{E[j]}} \equiv \frac{\alpha}{\beta}
\end{equation}
for real numbers $\alpha$ and $\beta$ in which case we have
\begin{equation}
\frac{\mathsf{E[i]}}{\mathsf{E[j]}} \equiv \frac{\text{Prob}(\mathsf{E[i]})}{\text{Prob}(\mathsf{E[j]})}
\end{equation}
These equivalence relations, (\ref{probratioruleequiv}) in particular, illustrate, once again, that there is a striking similarity between the form of the duotetensorial calculation and the form of the operational description.  If we extend our notion of equivalence to ratios in this way then this similarity applies to general fragments as well as circuits.

\section{Uploading physical theories into the duotensor framework}

\subsection{General idea}

We can \lq\lq upload" a physical theory into the duotensor framework if it pertains to a physical situation that can be described operationally with operations and wires and it satisfies Assumptions 1 and 2.  To do this we need
\begin{enumerate}
\item A choice of fiducial effects and preparations for each system type.
\item An expression for the fiducial probabilities for each possible operation (these are the proabilities with fiducial preparations on the inputs and fiducial effects on the outputs).  This gives us the duotensor with all black dots (as illustrated in (\ref{allblackduotensor})).
\item An expression for the hopping metric $\bbdots$ for each system type. The entries in this are the probabilities of the fiducial preparations followed by the fiducial effects.  We can invert $\bbdots$ to get $\wwdots$.
\end{enumerate}
Once we have the duotensor in one form (all black dots in this case) we can change the colour of the dots using $\wwdots$ and $\bbdots$ and fit the duotensors together to form duotensorial expressions that are equivalent to circuits and fragments.  In this way we can calculate probabilities for circuits and, for fragments, say whether probability ratios are well conditioned and what they are equal to.

\subsection{Uploading classical probability theory}

Classical probability theory concerns systems such as coins and dice which have a let of underlying states the system can be in.  A system $\mathsf a$ has $N_\mathsf a$ underlying states.  For example, for a coin $N_\mathsf{c}=2$ and for a die $N_\mathsf{d}=6$.  A preparation, $\mathsf C^{\mathsf{a}_1}$, will put the system in a state that is a probabilistic mixture of the underlying states:
\begin{equation}
{\bf p}(\mathsf{C}^{\mathsf{a}_1}) = \left( \begin{array}{c} p_1 \\ p_2 \\ \vdots \\ p_{N_\mathsf a} \end{array} \right)
\end{equation}
For composite systems
\begin{equation}
N_\mathsf{ab}=N_\mathsf{a} N_\mathsf{b}
\end{equation}
and the state is, correspondingly, given by a list of $N_\mathsf{ab}$ probabilities.  If, at some given time, we have separate preparations then we can take the tensor product of the states.

An effect, $\mathsf{ D_{a_1}}$, is associated with a dual vector, ${\bf r}(\mathsf{D_{a_1}})$, such that
\begin{equation}
\text{Prob}= {\bf r}\cdot {\bf p}
\end{equation}
If we have separate effects at a given time then we take the tensor product of the corresponding ${\bf r}$ vectors.

An operation $\mathsf{A_{a_1}^{b_2}}$ is associated with a $N_\mathsf{b}\times N_\mathsf{a}$ transformation matrix
\begin{equation}
Z(\mathsf{A_{a_1}^{b_2}})
\end{equation}
having entries (i) that are real, (ii) that are non-negative, and (iii) whose sum for each column is less than or equal to one.   In general
\begin{equation}
\text{Prob}= {\bf r}\cdot Z {\bf p}
\end{equation}
Note that for a set of operations $\mathsf{A_{a_1}^{b_2}}[i]$ having the same knob setting but different outcome sets $\mathsf{o_i}$ which are disjoint and sum to the the set of all outcomes, $\mathsf{o_I}$, we have that
\begin{equation}
\sum_i Z(\mathsf{A_{a_1}^{b_2}[i]}) = Z(\mathsf{A_{a_1}^{b_2}[I]})
\end{equation}
and, further $Z(\mathsf{A_{a_1}^{b_2}[I]})$ is a stochastic matrix (its entries are all non-negative and its columns sum to one).  We might be tempted to normalize each of the $Z(\mathsf{A_{a_1}^{b_2}[i]})$ matrices so they are also stochastic but then we would lose important information.  By neither normalizing these matrices or the state associated with the preparation  we are able to calculate the {\it joint} probability for the outcomes on the preparation $\mathsf C$, the operation $\mathsf A$ and the effect $\mathsf D$ (i.e.\ the probability for the given circuit).  If we had normalized the state and the transformation matrix we would obtain the probability for the outcome on the effect, $\mathsf D$, {\it conditioned} on seeing the given outcomes for the preparation $\mathsf C$ and the operation $\mathsf A$.  If we want to calculate conditional probabilities we can do so using Bayes rule afterwards rather than modifying the elements of the theory in the middle of a calculation.

To put this into the duotensor framework we
\begin{enumerate}
\item Pick a set of linearly independent states ${\bf p}({}_{a_1}\!X^{\mathsf{a}_1})$ for the fiducial preparations and a set of linearly independent effects ${\bf r}(X_{\mathsf{a}_1}^{a_1})$ for each system type.
\item Calculate the fiducial probabilities for each type of operation, $\mathsf{A_{a_1b_2\dots c_3}^{d_4e_5\dots f_6}}$, using
\begin{equation}
{\bf r}(\mathsf{X}{}_{\mathsf{d_4\, e_5 \, \dots \, f_6}}^{d_4 e_5 \dots f_6})
\cdot Z(\mathsf{ A_{a_1b_2\dots c_3}^{d_4e_5\dots f_6}}) 
{\bf p}(\mathsf{X}_{a_1 b_2\dots c_3}^{\mathsf{a_1 b_2\dots c_3}})
\end{equation}
where
\begin{equation}
{\bf r}(\mathsf{X}{}_{\mathsf{d_4\, e_5 \, \dots \, f_6}}^{d_4 e_5 \dots f_6}) := 
{\bf r}(X_{\mathsf{d}_4}^{d_4}) \otimes {\bf r}(X_{\mathsf{e}_5}^{e_5})\otimes \dots \otimes{\bf r}(X_{\mathsf{f}_6}^{f_6})
\end{equation}
and
\begin{equation}
{\bf p}(\mathsf{X}_{a_1 b_2\dots c_3}^{\mathsf{a_1 b_2\dots c_3}}) := 
{\bf p}({}_{a_1}\! X^\mathsf{a_1}) \otimes {\bf p}({}_{b_2}\! X^\mathsf{b_2})\otimes\dots \otimes {\bf p}({}_{c_3}\! X^\mathsf{c_3})
\end{equation}
This is equal to the duotensor with all black dots
\begin{equation}
{}_{a_1 b_2 \dots c_3}\!A^{d_4 e_5 \dots f_6}
\Longleftrightarrow
\begin{Diagram}{0}{0}
\Duobox[5]{A}{0,0}\putlatex[-45,-10]{\ensuremath{\vdots}}\putlatex[45,-10]{\ensuremath{\vdots}}
\inblack{A}{1}\Duosymbol{a}
\inblack{A}{2}\Duosymbol{b}
\inblack{A}{5}\Duosymbol{c}
\outblack{A}{1}\Duosymbol[0,4]{d}
\outblack{A}{2}\Duosymbol{e}
\outblack{A}{5}\Duosymbol{f}
\end{Diagram}
\end{equation}
\item Calulate the hopping metric $\bbdots$
\begin{equation}
{}_{a_1}\! g ^{a_1} = {\bf r}(X_{\mathsf{a}_1}^{a_1}) \cdot {\bf p}({}_{a_1}\!X^{\mathsf{a}_1})
\end{equation}
and $\wwdots$ by taking the inverse of this.
\end{enumerate}
It is worth making a few comments here.  The usual formulation of classical probability theory in terms of ${\bf p}$, $\bf r$, and $Z$ is pretty close to the duotensor formulation.  In particular, it corresponds to making the special choice of fiducial preparations corresponding to preparation of the underlying states (these correspond to $\bf p$ vectors having a 1 in the corresponding position and 0's elsewhere) and the special choice of effects corresponding to looking to see if the system is in the underlying states (these correspond to ${\bf r}$ vectors having a 1 in the corresponding position and 0's elsewhere).  For such a choice the hopping metric, $\bbdots$, is equal to the identity and therefore  so is its inverse $\wwdots$.  Hence changing the colour of the dots has no effect on the duotensor.  While this simplifies some real calculations, it hides much of the mathematical structure of the situation.  In particular, it hides structure that is essential in quantum theory.

\subsection{Uploading quantum theory into the duotensor framework}

A quantum system is associated with a complex Hilbert space, ${\mathcal H}_{N_\mathsf{a}}$, of dimension $N_\mathsf{a}$ that depends on the type of system.  A composite system $\mathsf{ab}$ is associated with a Hilbert space of dimension $N_\mathsf{ab}=N_\mathsf{a} N_\mathsf{b}$.

Let $\mathcal{V}_{N_\mathsf{a}}$ be the space of Hermitian operators that act on this.  All positive operators are Hermitean.
A preparation, $\mathsf{C^{a_1}}$, is associated with a positive operator,  $\hat{P}(\mathsf{C^{a_1}})\in\mathcal{V}_{N_\mathsf{a}}$. This positive operator must have trace less than or equal to one.
An effect, $\mathsf{D_{a_1}}$, is also associated with a positive operator, $\hat{ P}(\mathsf{D_{a_1}})\in\mathcal{ V}_{N_\mathsf{a}}$.  This positive operator must have the property that $\hat{I} - \hat{P}$ is also a positive (or zero) operator where $I$ is the identity.

An operation, $\mathsf{A_{a_1}^{b_2}}$, is associated with a superoperator, $\$(\mathsf{A_{a_1}^{b_2}})$, that acts on operators in $\mathcal{ V}_{N_\mathsf{a}}$ and returns operators in $\mathcal{ V}_{N_\mathsf{b}}$.  The superoperator must be completely positive (this means that when $\$\otimes I$ acts on any positive operator in $\mathcal{V}_{N_\mathsf{a}}\otimes{\mathcal V}_{N_\mathsf{c}}$ it returns a positive operator in
$\mathcal{V}_{N_\mathsf{b}}\otimes\mathcal{V}_{N_\mathsf{c}}$ where $I$ is the identity acting on ${\mathcal V}_{N_\mathsf{c}}$ for any $N_\mathsf{c}$).  It must also be completely trace non-increasing (this means that when $\$\otimes I$ acts on any positive operator it returns a positive operator having trace that is less than or equal to that of the original positive operator).  Preparations and effects are special cases of superoperators having the trivial system as input and output respectively.

The probability for the circuit $\mathsf{C^{a_1}A_{a_1}^{b_1}E_{b_1}}$ is
\begin{equation}
\text{Prob}(\mathsf{C^{a_1}A_{a_1}^{b_1}E_{b_1}})= \text{Trace}\left(\hat{P}(\mathsf{E_{b_1}}) \$(\mathsf{A_{a_1}^{b_2}}) \hat{P}(\mathsf{C^{a_1}})\right)
\end{equation}
Note that this is the joint probability for seeing the outcomes at the preparation $\mathsf C$, the operation $\mathsf A$, and the effect $\mathsf E$ rather than the conditional probability of seeing the outcome for $\mathsf E$  given that we have seen the outcomess for $\mathsf C$ and $\mathsf A$. This is the reason we allow positive operators representing preparations to have trace less than one, and why we only demand that the superoperators is trace non-increasing (it can be trace decreasing).

We are now in a position to put quantum theory in the duotensor framework.
\begin{enumerate}
\item  For the fiducial preparations, ${}_{a_1}\!\mathsf{X}^{\mathsf{a}_1}$, we chose any set of $N_\mathsf{a}^2$ linearly independent positive operators (having trace less than or equal to one), $\hat{P}({}_{a_1}\!\mathsf{X}^{\mathsf{a}_1})$. Here $a_1= 1$ to $K_\mathsf{a}= N_\mathsf{a}^2$. These span the space $\mathcal{V}_{N_\mathsf{a}}$.  We do this for each system type.

    For the fiducial effects, $\mathsf{X}_{\mathsf{a}_1}^{a_1}$, we choose any set of linearly independent positive operators $\hat{P}(\mathsf{X}_{\mathsf{a}_1}^{a_1})$ (that are such that $\hat{I}-\hat{P}$ is also positive).  Here $a_1= 1$ to $K_\mathsf{a}= N_\mathsf{a}^2$. These span the space $\mathcal{V}_{N_\mathsf{a}}$.  We do this for each system type.
\item The fiducial probabilities for each operation $\mathsf{A_{a_1b_2\dots c_3}^{d_4e_5\dots f_6}}$ are given by
\begin{equation}
\text{Trace}\left[ \hat{P}(\mathsf{X}{}_{\mathsf{d_4 \, e_5 \, \dots \, f_6}}^{d_4 e_5 \dots f_6})
\$ (\mathsf{A_{a_1b_2\dots c_3}^{d_4e_5\dots f_6}})
\hat{P}(\mathsf{X}_{a_1 b_2\dots c_3}^{\mathsf{a_1 b_2\dots c_3}})
\right]
\end{equation}
where
\begin{equation}
\hat{P}(\mathsf{X}{}_{\mathsf{d_4\, e_5 \, \dots \, f_6}}^{d_4 e_5 \dots f_6}) := \hat{P}(\mathsf{X}_{\mathsf{d}_4}^{d_4})\otimes \hat{P}(\mathsf{X}_{\mathsf{e}_5}^{e_5})\otimes \dots \otimes
\hat{P}(\mathsf{X}_{\mathsf{f}_6}^{f_6})
\end{equation}
and 
\begin{equation}
\hat{P}(\mathsf{X}_{a_1 b_2\dots c_3}^{\mathsf{a_1 b_2\dots c_3}}) := \hat{P}({}_{a_1}\!\mathsf{X}^{\mathsf{a}_1}) \otimes \hat{P}({}_{b_2}\!\mathsf{X}^{\mathsf{b}_2})\otimes \dots
\otimes \hat{P}({}_{c_3}\!\mathsf{X}^{\mathsf{c}_3})
\end{equation}
This is equal to the duotensor with all black dots
\begin{equation}
{}_{a_1 b_2 \dots c_3}\!A^{d_4 e_5 \dots f_6}
\Longleftrightarrow
\begin{Diagram}{0}{0}
\Duobox[5]{A}{0,0}\putlatex[-45,-10]{\ensuremath{\vdots}}\putlatex[45,-10]{\ensuremath{\vdots}}
\inblack{A}{1}\Duosymbol{a}
\inblack{A}{2}\Duosymbol{b}
\inblack{A}{5}\Duosymbol{c}
\outblack{A}{1}\Duosymbol[0,4]{d}
\outblack{A}{2}\Duosymbol{e}
\outblack{A}{5}\Duosymbol{f}
\end{Diagram}
\end{equation}
\item We calulate the hopping metric $\bbdots$
\begin{equation}
{}_{a_1}\! g ^{a_1} = \text{Trace}\left( \hat{P}(\mathsf{X}_{\mathsf{a}_1}^{a_1})\hat{P}({}_{a_1}\!\mathsf{X}^{\mathsf{a}_1})\right)
\end{equation}
By taking the inverse of this we get $\wwdots$.  We do this for each system type.
\end{enumerate}
The duotensor $\wwdots$ will have negative entries.  This is where the negative numbers come from in real number representations of quantum theory.

\section{Foliations}

In \cite{foliable} the author showed how to formulate general probabilistic theories with definite causal structure with respect to arbitrary foliations. In that formulation it was necessary to foliate a circuit with spacelike hypersurfaces in order to calculate the probability for the circuit.  In the present framework this is not necessary.  We can cut a circuit up into arbitrary fragments which need not correspond to anything like spacelike slices. It is interesting, nevertheless, to see how we might go about foliating a circuit in the present framework and using this foliation to calculate the probability for the circuit.  Consider the circuit
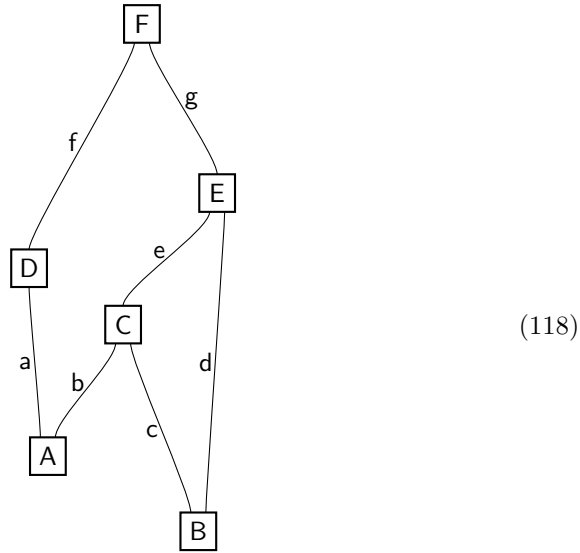
\begin{equation}\label{mediumcircuit}
\begin{Diagram}{0}{-1.8}
\Opbox[2]{A}{0,0} \Opbox[2]{B}{8,-4}
\Opbox[2]{D}{-1,10}
\Opbox[2]{C}{4,7}
\Opbox[2]{E}{9,14}
\Opbox[2]{F}{5,23}
\wire{A}{D}{1}{1.5} \opsymbol{a} \wire{A}{C}{2}{1} \opsymbol[0,5]{b} \wire{B}{C}{1}{2} \opsymbol{c} \wire{B}{E}{2}{2}\opsymbol{d}
\wire{C}{E}{1.5}{1}\opsymbol{e} \wire{D}{F}{1.5}{1} \opsymbol{f} \wire{E}{F}{1.5}{2} \otherside\opsymbol[0,5]{g}
\end{Diagram}
\end{equation}
We can foliate this circuit in many different ways. One way is the following
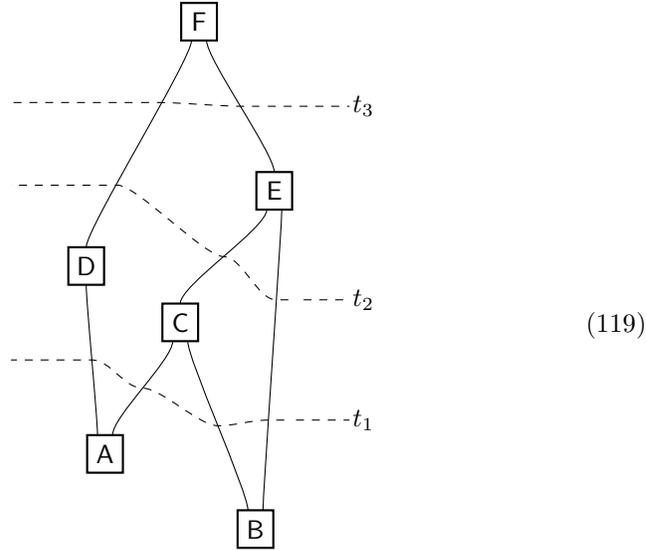
\begin{equation}
\begin{Diagram}{0}{-1.8}
\Opbox[2]{A}{0,0}
\Opbox[2]{B}{8,-4}
\Opbox[2]{D}{-1,10}
\Opbox[2]{C}{4,7}
\Opbox[2]{E}{9,14}
\Opbox[2]{F}{5,23}
\wire{A}{D}{1}{1.5} \wire{A}{C}{2}{1}  \wire{B}{C}{1}{2}  \wire{B}{E}{2}{2}
\wire{C}{E}{1.5}{1}\wire{D}{F}{1.5}{1}  \wire{E}{F}{1.5}{2}
\begin{foliation}{-5}{13}
\startfoliate{A}{D}{1}{1.5}\continuefoliate{A}{C}{2}{1}\continuefoliate{B}{C}{1}{2}\Finishfoliate{B}{E}{2}{2}{-0.2} \otherside\putlatex{\ensuremath{t_1}}
\Startfoliate{D}{F}{1.5}{1}{-0.2}\continuefoliate{C}{E}{1.5}{1}\Finishfoliate{B}{E}{2}{2}{0.2}
\otherside\putlatex{\ensuremath{t_2}}
\Startfoliate{D}{F}{1.5}{1}{0.2}\finishfoliate{E}{F}{1.5}{2}
\otherside\putlatex{\ensuremath{t_3}}
\end{foliation}
\end{Diagram}
\end{equation}
The hypersurfaces in this foliation consist of sets of wires (those wires intersected by the dashed line).  These wires form a synchronous set (it is not possible to reach any wire in the set from any other wire by tracing forward along wires from output to input through the circuit). The foliation shown in this diagram is complete.  By this we mean that every wire is included in at least one hypersurface.  This circuit is equivalent to the duotensorial calculation
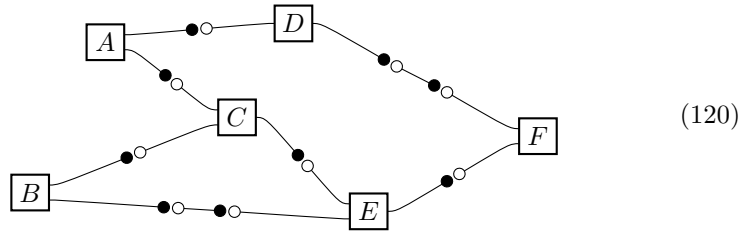
\begin{equation}
\begin{Diagram}{0}{0.9}
\Duobox[2]{A}{0,0}
\Duobox[2]{B}{-4,-8}
\Duobox[2]{D}{10,1}
\Duobox[2]{C}{7,-4}
\Duobox[2]{E}{14,-9}
\Duobox[2]{F}{23,-5}
\linkbw{A}{D}{1}{1.5} \linkbw{A}{C}{2}{1}  \linkbw{B}{C}{1}{2}  \linkbwbw{B}{E}{2}{2}
\linkbw{C}{E}{1.5}{1}\linkbwbw{D}{F}{1.5}{1} \linkbw{E}{F}{1.5}{2}
\end{Diagram}
\end{equation}
which we can break up into four duotensors
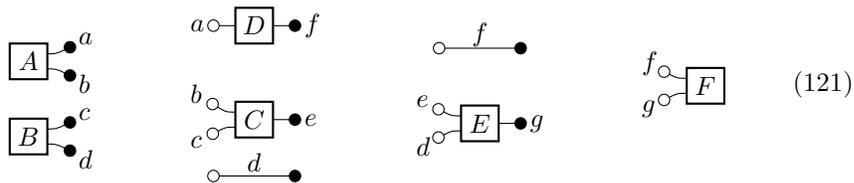
\begin{equation}
\begin{Diagram}{0}{0.9}
\begin{move}{-4,-0.7}
\Duobox[2]{A}{0,-2}\outblack[-5]{A}{1}\Duosymbol{a} \outblack[5]{A}{2} \Duosymbol{b}
\Duobox[2]{B}{0,-6} \outblack[-5]{B}{1}\Duosymbol{c} \outblack[5]{B}{2} \Duosymbol{d}
\end{move}
\begin{move}{-2,-1.8}
\Duobox[2]{D}{10,1}\inwhite[0]{D}{1.5}\Duosymbol{a} \outblack[0]{D}{1.5}\Duosymbol{f}
\Duobox[2]{C}{10,-4}\inwhite[-6]{C}{1} \Duosymbol{b} \inwhite[5]{C}{2} \Duosymbol{c} \outblack{C}{1.5}\Duosymbol{e}
\wbmetric[1.5]{g1}{10,-7}\duosymbol[0,5]{d}
\end{move}
\begin{move}{0,0}
\Duobox[2]{E}{20,-6}\inwhite[-5]{E}{1}\Duosymbol{e} \inwhite[5]{E}{2}\Duosymbol{d} \outblack[0]{E}{1.5}\Duosymbol{g}
\wbmetric[1.5]{g2}{20,-2} \duosymbol[0,5]{f}
\end{move}
\begin{move}{2,1}
\Duobox[2]{F}{30,-5} \inwhite[-5]{F}{1}\Duosymbol{f}  \inwhite[5]{F}{2}\Duosymbol{g}
\end{move}
\end{Diagram}
\end{equation}
The first duotensor corresponds to the preparation $\mathsf{AB}$ and it can be thought of as providing the state at time $t_1$. The preparation $\mathsf{AB}$ is a product of two preparations, $\mathsf A$ and $\mathsf B$, so the state is a product state between the $\mathsf{ab}$ and $\mathsf{cd}$ systems.  The next duotensor can be thought of as a transformation matrix that acts on the state at time $t_1$ to give the new state at time $t_2$.  This transformation acts as the identity on system $\mathsf d$ while acting non-trivialy on system $\mathsf{abc}$.  The third duotensor transforms the state from time $t_2$ to time $t_3$. The final duotensor corresponds to an effect.  We have deliberately written put black dots on outputs and white dots on inputs.  This ensures that the state is always a list of probabilities.  Of course, this is not actually necessary but corresponds to the usual choice in operational frameworks.   Any circuit can be given a complete foliation (so long as there are no closed loops) and so we can always calculate the probability associated with a circuit by thinking in terms of a state evolving in time. However, viewed from the point of view of the duotensor framework, it is deeply unnatural to insist on foliating the circuit to calculate its probability.  We can just as well break it up into fragments of any other sort.  Further more, if we do insist on foliating like this then, in many cases, we will need to pad the calculation with identities (i.e. the $\wbdots$).  There are two identities in the example we just considered.  In fact any complete foliation of the circuit in (\ref{mediumcircuit}) will have some such padding (this is clearly a fairly generic property of foliations of circuits).  If we do not insist on foliating the circuit then there are many other ways of breaking the circuit up into fragments which will not lead to such padding.  Indeed, we could simply drop the identities:
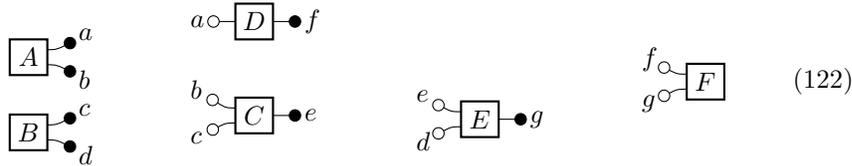
\begin{equation}
\begin{Diagram}{0}{0.9}
\begin{move}{-4,-0.7}
\Duobox[2]{A}{0,-2}\outblack[-5]{A}{1}\Duosymbol{a} \outblack[5]{A}{2} \Duosymbol{b}
\Duobox[2]{B}{0,-6} \outblack[-5]{B}{1}\Duosymbol{c} \outblack[5]{B}{2} \Duosymbol{d}
\end{move}
\begin{move}{-2,-1.8}
\Duobox[2]{D}{10,1}\inwhite[0]{D}{1.5}\Duosymbol{a} \outblack[0]{D}{1.5}\Duosymbol{f}
\Duobox[2]{C}{10,-4}\inwhite[-6]{C}{1} \Duosymbol{b} \inwhite[5]{C}{2} \Duosymbol{c} \outblack{C}{1.5}\Duosymbol{e}
\end{move}
\begin{move}{0,0}
\Duobox[2]{E}{20,-6}\inwhite[-5]{E}{1}\Duosymbol{e} \inwhite[5]{E}{2}\Duosymbol{d} \outblack[0]{E}{1.5}\Duosymbol{g}
\end{move}
\begin{move}{2,1}
\Duobox[2]{F}{30,-5} \inwhite[-5]{F}{1}\Duosymbol{f}  \inwhite[5]{F}{2}\Duosymbol{g}
\end{move}
\end{Diagram}
\end{equation}
Here we have another four duotensors. We can sum over these duotensors to give the probability for the circuit in (\ref{mediumcircuit}) as follows:
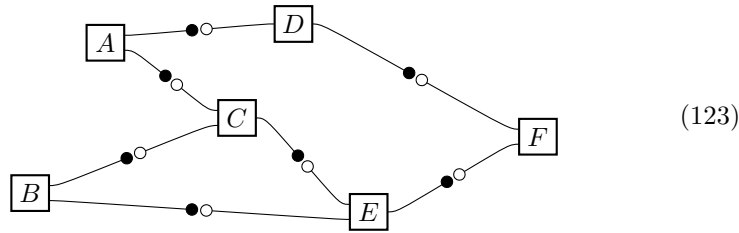
\begin{equation}
\begin{Diagram}{0}{0.9}
\Duobox[2]{A}{0,0}
\Duobox[2]{B}{-4,-8}
\Duobox[2]{D}{10,1}
\Duobox[2]{C}{7,-4}
\Duobox[2]{E}{14,-9}
\Duobox[2]{F}{23,-5}
\linkbw{A}{D}{1}{1.5} \linkbw{A}{C}{2}{1}  \linkbw{B}{C}{1}{2}  \linkbw{B}{E}{2}{2}
\linkbw{C}{E}{1.5}{1}\linkbw{D}{F}{1.5}{1} \linkbw{E}{F}{1.5}{2}
\end{Diagram}
\end{equation}
As simple examination shows, this does not correspond to a foliation. Furthermore, it is a simpler calculation than that corresponding to the foliation we considered.

The foliation point of view is, in fact, an advance on the more common Newtonian view of circuits.  In the Newtonian view it is assumed that there is a background time and each operation happens over some given time interval with respect to this background time.  In this case the circuits cannot be interpreted graphically since moving a box up or down on the page will change the physical interpretation.  However, both the Newtonian and the foliation points of view are based on a picture of physics in which we make predictions by evolving a state with respect to some time coordinate.  Despite being deeply ingrained in the way we think about the world, it is not necessary to think in terms of an evolving state.   We can instead cut the world up into arbitrary shaped fragments and then calculate probabilities by putting these bits together as in the duotensor framework.  It is clear that a framework which depends on providing a foliation will not have the formalism locality property.  The duotensor framework has the formalism locality property because it allows us to consider partitionings into arbitrary shaped fragments.

\section{Related work}\label{relatedwork}

This paper brings together separate strands of work.  It combines ideas from the authors papers \cite{fiveaxioms, causaloid1, foliable} and is much influenced by the use of graphical calculus in quantum theory initiated by Abramsky and Coecke \cite{AbramskyCoecke} (see also \cite{quantumpicturalism}; see \cite{Selinger} for a general review of graphical languages of relevance to this type of physical situation).

In \cite{fiveaxioms} a general probabilistic framework (sometimes called the $\bf r$-$\bf p$ framework) is developed for the purpose of deriving quantum theory from simple axioms.  The $\bf r$-$\bf p$ framework is actually a simple example of a framework that has been developed over the years by many authors \cite{Mackey, Ludwig, DaviesLewis, Wootters1, FoulisRandall, Gudder, Barrett, CDP, BarnumWilce} and is sometimes called the \lq\lq convex probabilities approach".  In this framework states are represented vectors whose entries are probabilities for a fiducial set of effects.  Effects are represented by vectors in the dual space and transformations by matrices having real entries.  This framework uses the idea of an evolving state and so does not have the formalism locality property.

In \cite{causaloid1} (see also \cite{causaloid2, causaloid3, causaloid4}) the causaloid framework was developed for situations where we may have indefinite causal structure (with the hope of providing a route to a theory of quantum gravity).  This framework does have the formalism locality property.   The framework presented in this paper, although less general than the causaloid framework, is much simpler because it makes use of the input/output structure of operations.  It is much easier to see how to put classical probability theory and quantum theory into this framework.

In the convex probabilities approach a rather Newtonian attitude is usually taken towards time.  Time is a background parameter.  The vertical position of an operation on the page (within a circuit) corresponds to the time at which the operation happens.  In \cite{foliable} a different attitude is taken.  Only the graphical information in the circuit has meaning.  Thus, we can move the operation up and down on the page without changing the physical interpretation of the circuit so long as we do not change the wiring.  Circuits are analysed by \lq\lq foliating" to reintroduce a time parameter with respect to which we can evolve a state.  For quantum theory this approach goes back to Tomonaga and Schwinger \cite{TS, TS2}.  In the graphical context Markopoulou \cite{Fotini} put forward the quantum causal histories framework.  In this approach processes are associated with edges and systems with vertices.  Blute, Ivanov, and Panangaden \cite{BIP} give a dual formulation in which processes are associated with vertices and systems with edges and it is this dual attitude that is adopted in \cite{foliable}. The approach of foliating circuits for the purpose of doing quantum field theory has been studied recently by D'Ariano \cite{DAriano}.  We are unable to have the formalism locality property if must foliate a circuit to do a calculation.  Furthermore, the foliation approach necessitates padding calculations with the identity (since sometimes more than one hypersurface in the foliation cuts a given wire).  The technique in this paper does not require that we foliate and consequently we do not need to pad with the identity in this way.

The notion of a fragment or something quite similar appears in various guises in many papers.  Aharanov and collaborators have have studied {\it pre- and post-selected quantum ensembles} \cite{ABL} (and more complicated situations \cite{Aharonovetal}) which clearly are similar to fragments.  Oeckl \cite{Oeckl} has given a {\it general boundary} formulation of quantum theory developed by Segal \cite{Segal} and Atiyah \cite{Atiyah} in which spacetime can be divided up into arbitrary spacetime regions then glued back together. This is related to topological quantum field theory in which something similar can be done.  Quantum theory was formulated in the causaloid framework \cite{causaloid1} allowing arbitrary spacetime regions.    Gutoski and Watrous \cite{strategies} define {\it strategies} which are basically the same as fragments for quantum circuits and are of use in studying quantum protocols.   Chiribella, D'Ariano, and Perinotti \cite{combs} invented {\it quantum combs} which are also similar to fragments. The \lq\lq link product" in their paper provides a way of putting together quantum maps to produce a map for the comb and is clearly related to duotensor composition in this paper.  The advantage of the duotensor approach is that it is formulated for general probabilistic theories (rather than just quantum theory), it has the formalism locality property, and it is conceptually and mathematically very simple.

Abramsky and Coecke \cite{AbramskyCoecke} (see also \cite{quantumpicturalism, EricBob} and references therein) showed how to formulate quantum theory in category theoretic terms.  This brought with it the use of pictures to prove theorems.  Pictures often furnish much more immediate proofs than symbolic notation does.  The diagrammatic notation in this paper is not the same as that of Abramsky and Coecke but is clearly inspired by that work.  We do not engage in category theoretic analysis though such an analysis would probably be very illuminating.  Chiribella, D'Ariano, Perinotti have also adopted a pictorial approach in their recent paper \cite{CDP} providing a general framework for probabilistic theories.

\section{Discussion}

Assumption 2 is the key new assumption here.  It it a hybrid statement.  It provides a link between the physics (the operational level of description) and the mathematics (the duotensors) of the associated calculation.  This assumption enables us to translate between a diagram (or symbolic expression) describing the physics to one describing the mathematicians.  We could search for assumptions expressed as hybrid statements that enable a similar translation between physics and mathematics for other types of physical theory (for example, General Relativity).  Indeed this suggests the following principle:
\begin{quote}
{\bf Physics to mathematics correspondence principle.} For any physical theory, there exists a small number of simple hybrid statement that enable us to translate from the physical description to the corresponding mathematical calculation such that the mathematical calculation (in appropriate notation) looks the same as the physical description (in appropriate notation).
\end{quote}
Such a principle might be useful in obtaining new physical theories (such as a theory of quantum gravity).  Related ideas to this have been considered by category theorists \cite{EricBob}.  A category of physical processes can be defined corresponding to the physical description.  A category corresponding to the mathematical calculation can also be given.  The mapping from the first category to the second is given by a functor (this takes us from one category to another).

A standard assumption made in operational frameworks is the assumption that local tomography is possible \cite{Araki, Bergia, Wootters86, Mermin, HardyWootters}.  This is the assumption that the state of a composite system can be determined from the joint probabilities obtained by making separate measurements on the components.  Assumption 2 implies the local tomography principle (we can determine the duotensor for a preparation from the fiducial probabilities obtained by putting fiducial effects on each of the outputs of the preparation) and hence is a stronger assumption. However, in the context of the circuit model, Assumptions 1 and 2 are enough to obtain the basic framework whereas, when local tomography is used, an additional causality type assumption must be made \cite{CDP,foliable}.

We have obtained the duotensor formalism under the assumption that circuits are possible.  This requires that there exist operations with no inputs and operations with no outputs.  It may be that all operations have both inputs and outputs.  In such a case we can still use the duotensor formalism to do calculations as long as we know the duotensor associated with each operation. Further, it is possible that such calculations would actually yield the correct answers. We can, then, formulate physical theories without having the possibility of closed circuits.  In such a case it would be good to look for different assumptions to the two used in this paper to motivate the formalism.

The hopping metric $\bbdots$ causes indices to hop from left to right (or vice versa).  It may be interesting to consider jumping metrics $\bbleftjump$ and $\bbrightjump$ which cause indices to be raised (jump up) and lowered (jump down).  In fact it would be interesting to define the full range of such objects
\begin{equation}
\bbleftjump ~~~~ \bwleftjump ~~~~ \wbleftjump ~~~~ \wwleftjump ~~~~\bbrightjump ~~~~ \bwrightjump ~~~~ \wbrightjump ~~~~ \wwrightjump
\end{equation}
Such symbols may be related to the cups and caps used to great effect by Abramsky, Coecke, and collaborators \cite{AbramskyCoecke, quantumpicturalism, EricBob}.

The approach in this paper was for finite systems only.  By this we mean that (i) any operation or fragment has a finite number of inputs and outputs, and (ii) each system type is associated with a finite $K_\mathsf{a}$.  To do field theory for continuous fields we would need to relax these requirements.  One possible way to proceed is the following.  Work with a fixed Minkowski background.  Each fragment would be associated with a spacetime region having a boundary.  We are able to wire together two fragments if some part of their boundaries fit together (this may require a boost).  We can consider infinitesimal areas on the boundary.  Associated with each infinitesimal area with outward normal pointing to the future would be an output. Associated with each infinitesimal area with outward normal pointing to the past would be an input.  Associated with any part of the boundary with normal pointing in a spacelike direction would be both an input and an output.  The type associated with the input and output would be determined by (a) the type of field and (b) the invariant area of the infinitesimal. Under system composition the areas would add.   By imposing Assumptions 1 and 2 we could set up the duotensor framework so long as we are able to consistently replace sums with integrals as required.  If this worked it would allow us to do both classical field theory (for a probabilistic version of electromagnetism for example) and quantum field theory.  We could not do general relativity this way since we have assumed a fixed background metric.

The approach in this paper was motivated by considerations having to do with the construction of a theory of quantum gravity.  However, it seems unlikely that this approach could be directly applied to the problem of finding such a theory since it is within the circuit model which imposes a kind of definite causal structure coming from the wires.  Nevertheless, it is hoped that, by relaxing the circuit model, the ideas in this paper may find application to the problem of quantum gravity.


\section*{Acknowledgements}
Research at Perimeter Institute for Theoretical Physics is supported in part by
the Government of Canada through NSERC and by the Province of Ontario
through MRI.

\newpage

\appendix
\section{The duotenzor drawing package}

All the diagrams in this paper have been drawn using commands in the purpose built \verb+duotenzor+ package (spelled with a \verb+z+).  This is a purpose built package for drawing circuits and duotensor diagrams.  It consists of about eighty commands (defined using the LaTeX \verb+\newcommand+ command) that call on the TikZ package written by Till Tantau \cite{TikZ}. TikZ is, of course, much much more versatile and it (or some other similar package) should be used to draw more complicated diagrams.  \verb+duotenzor+ has two advantages:  (1) fewer commands are required for circuit and duotensor diagrams than with a more advanced package; (2) producing a drawing is very similar to producing an equation using LaTeX.  The package file, \verb+duotenzor.sty+, can be downloaded from CTAN \cite{CTAN} and, once saved into the appropriate folder (the working folder will suffice), uploaded into the document using the command \verb+\usepackage{duotenzor}+ in the document preamble.  This package calls on the TikZ drawing package \cite{TikZ} and two of its libraries (\verb+calc+ and \verb+arrows+).  It also calls on the \verb+xspace+ package.  All these are included as standard in most LaTeX installations. 

\verb+duotenzor+ consists of about sixty commands which must be used inside
\verb+\begin{diagram} ... \end{diagram}+ commands.  In addition there are about twenty commands that can be used in regular text or in math mode but not inside the diagram environment.  The following code produces the example on the right.

\begin{verbatim}
\begin{diagram}
\Opbox{A}{0,0}
\Opbox{B}{2,4}
\wire{A}{B}{1}{2}
\end{diagram}
\end{verbatim}
\makebox[3cm][l]{
\begin{diagram}
\boundingbox{0,0}{0,0}
\begin{move}{30,5}
\Opbox{A}{0,0}
\Opbox{B}{2,4}
\wire{A}{B}{1}{2}
\end{move}
\end{diagram}
}

\vspace{-13pt}
\noindent
Here \verb+\Opbox{B}{2,4}+ puts a box at coordinate $(2,4)$ with the symbol $\mathsf B$ in it.  An alternative way to do this is to use
\verb+\opbox{B}{2,4}\opsymbol{B}+.  In this case \verb+\opbox{B}{2,4}+ produces an empty box which, in the code, is called \verb+B+ (we cannot have two objects with the same name).  Then the command \verb+\opsymbol{B}+ puts a $\mathsf B$ in the box.  We need to use this more lengthy construction if we want to have two boxes with the same letter inside them.  By default these boxes are of size 3 having room for 3 inputs and 3 outputs.  A box of size 5 is produced by \verb+\Opbox[5]{A}{2,4}+.  We can choose non-integer sizes should we wish.  The \verb+\wire{A}{B}{1}{2}+ command draws a wire from output 1 of box \verb+A+ to input 2 of box \verb+B+.  The inputs and outputs need not be integers.  This is particularly useful if we want to have a different number inputs and outputs.

The command \verb+\opsymbol{s}+ puts the symbol $\mathsf{s}$ (in the \verb+\mathsf+ font) at the appropriate location for the preceeding object.  We can use to put a symbol in a box or on the left of a wire for example. To switch the symbol to the other side of the wire we use the command \verb+\otherside+ immediately before the \verb+\opsymbol+ command.   The command allows optional fine tuning.  For example, \verb+\opsymbol[5,13]{s}+
will move the symbol 5 small units to the left and 13 small units up.  This fine tuning is useful since \verb+duotenzor+ does not always position the labels to ones taste.  Here is an example illustrating these points.

\begin{verbatim}
\begin{diagram}
\Opbox{A}{0,0}
\opbox[2]{B}{2,4}\opsymbol{B}
\Opbox[2]{C}{-1,9}
\wire{A}{B}{2}{1}\opsymbol{a}
\wire{A}{B}{3}{2} \otherside \opsymbol{c}
\wire{A}{C}{1}{1}\opsymbol{b}
\wire{B}{C}{1.5}{2} \opsymbol[-2,-5]{d}
\end{diagram}
\end{verbatim}
\makebox[3cm][l]{
\begin{diagram}
\boundingbox{0,0}{0,0}
\begin{move}{38,6.5}
\Opbox{A}{0,0}
\opbox[2]{B}{2,4}\opsymbol{B}
\Opbox[2]{C}{-1,9}
\wire{A}{B}{2}{1}\opsymbol{a}
\wire{A}{B}{3}{2} \otherside \opsymbol{c}
\wire{A}{C}{1}{1}\opsymbol{b}
\wire{B}{C}{1.5}{2} \opsymbol[-2,-5]{d}
\end{move}
\end{diagram}
}

\vspace{-13pt}
\noindent Here is another example illustrating the use of \verb+\inwire+ and \verb+\outwire+
\makebox[3cm][l]{
\begin{diagram}
\boundingbox{0,0}{0,0}
\begin{move}{-4,-9}
\Opbox{D}{0,0}
\inwire[-5]{D}{1} \Opsymbol{a}
\inwire{D}{2} \Opsymbol{b}
\closedinwire[5]{D}{3} \Opsymbol{c}
\outwire[-5]{D}{1.5} \Opsymbol{d}
\closedoutwire[5]{D}{2.5} \Opsymbol{e}
\end{move}
\end{diagram}
}
\begin{verbatim}
\begin{diagram}
\Opbox{D}{0,0}
\inwire[-5]{D}{1} \Opsymbol{a}
\inwire{D}{2} \Opsymbol{b}
\closedinwire[5]{D}{3} \Opsymbol{c}
\outwire[-5]{D}{1.5} \Opsymbol{d}
\closedoutwire[5]{D}{2.5} \Opsymbol{e}
\end{diagram}
\end{verbatim}

\noindent The optional argument in square brackets on \verb+\inwire+ and \verb+\outwire+ allows the inwire and outwire to be bent which can be visually more appealing.  We use \verb+\Opsymbol+ rather than \verb+\opsymbol+ for putting symbols at the end of wires to get the symbols aligned (\verb+\opsymbol+ can be used but produces messy results).  The commands \verb+\closedinwire+ and \verb+\closedoutwire+ produce closed inputs and outputs as shown.

Similar commands produce duotensor diagrams.  Here is an example
\makebox[3cm][l]{
\begin{diagram}
\boundingbox{0,0}{0,0}
\begin{move}{-9,-11}
\Duobox[2]{A}{0,0}
\duobox[2]{B}{6,3} \duosymbol{B}
\link{A}{B}{1}{1.5}\duosymbol{a}
\inblack[-5]{A}{1} \Duosymbol{a}
\inwhite[5]{A}{2} \Duosymbol{b}
\outblack[5]{A}{2} \duosymbol{d}
\outwhite[-5]{B}{1}\Duosymbol{a}
\outblack[5]{B}{2}\Duosymbol{c}
\end{move}
\end{diagram}
}
\begin{verbatim}
\begin{diagram}
\Duobox[2]{A}{0,0}
\duobox[2]{B}{6,3} \duosymbol{B}
\link{A}{B}{1}{1.5}\duosymbol{a}
\inblack[-5]{A}{1} \Duosymbol{a}
\inwhite[5]{A}{2} \Duosymbol{b}
\outblack[5]{A}{2} \duosymbol{d}
\outwhite[-5]{B}{1}\Duosymbol{a}
\outblack[5]{B}{2}\Duosymbol{c}
\end{diagram}
\end{verbatim}

The command \verb+\otherside+ can be used to put a label below (rather than above) a link.

\newpage

Often it is necessary to put black and white dots on the links themselves.  This is illustrated by
\makebox[3cm][l]{
\begin{diagram}
\boundingbox{0,0}{0,0}
\begin{move}{17,-11}
\Duobox[12]{A}{0,0}
\Duobox[12]{B}{12,4}
\linkbw{A}{B}{2}{2}
\linkwb{A}{B}{4}{4}
\linkwbbw{A}{B}{6}{6}
\linkbwwb{A}{B}{8}{8}
\linkwbwb{A}{B}{10}{10}
\linkbwbw{A}{B}{12}{12}
\end{move}
\end{diagram}
}
\begin{verbatim}
\begin{diagram}
\Duobox[12]{A}{0,0}
\Duobox[12]{B}{12,4}
\linkbw{A}{B}{2}{2}
\linkwb{A}{B}{4}{4}
\linkwbbw{A}{B}{6}{6}
\linkbwwb{A}{B}{8}{8}
\linkwbwb{A}{B}{10}{10}
\linkbwbw{A}{B}{12}{12}
\end{diagram}
\end{verbatim}

\noindent Both \verb+\wire+ and \verb+\link+ (along with all the related commands in the above example) also take an optional argument in square brackets which increases the curviness of the line.  The value 0 produces a straight line.  The default value is 1.

There are a few commands that are useful for drawing hybrid diagrams.  The command \verb+\Opduobox{hsize}{vsize}{A}{2,5}+ produces a box of horizontal size \verb+hsize+ and vertical size \verb+vsize+ labeled with $\mathsf A$ at coordinate $(2,5)$.  Both links and wires can be attached to this box.  The command \verb+\opduobox+ is the same except it does not have a symbol in it (a symbol can be placed in it using \verb+\opsymbol+ or \verb+\duosymbol+).  We also have \verb+\fideffect[scale]{F}{2,7}+ and \verb+\fidprep[scale]{G}{2,7}+ which produce triangles representing fiducial effects and preparations scaled by the optional \verb+scale+ argument.  They can be both linked to and wired to.  Here is an example
\makebox[3cm][l]{
\begin{diagram}
\boundingbox{0,0}{0,0}
\begin{move}{7,-12}
\Opduobox{4}{3}{A}{0,0}
\Opbox{B}{1,6}
\fideffect[2]{F}{-5,-2}\opsymbol{F}
\fidprep{G}{5,2}
\link{A}{G}{2}{1}
\link{F}{A}{1}{3}
\wire{A}{B}{4}{1}
\wire{G}{B}{1}{3}
\end{move}
\end{diagram}
}
\begin{verbatim}
\begin{diagram}
\Opduobox{4}{3}{A}{0,0}
\Opbox{B}{1,6}
\fideffect[2]{F}{-5,-2}\opsymbol{F}
\fidprep{G}{5,2}
\link{A}{G}{2}{1}
\link{F}{A}{1}{3}
\wire{A}{B}{4}{1}
\wire{G}{B}{1}{3}
\end{diagram}
\end{verbatim}

\noindent It is difficult to align the fiducial effects and fiducial preparations with the duobox they are linked to using the above commands. To help with this, the additional commands \verb+\linkedeffect+ and \verb+\linkedprep+ are supplied.  The command \verb+\linkedeffect[scale]{A}{3}{F}{x}{Dy}+ draws a fiducial element named \verb+F+ with $x$-coordinate \verb+x+ and a $y$-coordinate at a height \verb+Dy+ above the 3 port of duobox \verb+A+ (usually \verb+Dy+ is set to 0 since we are trying to align these objects).  Further, it draws in the link.  Here is an example
\newpage
\makebox[3cm][l]{
\begin{diagram}
\boundingbox{0,0}{0,0}
\begin{move}{35,-6}
\Duobox[2]{C}{0,0}
\linkedeffect[0.7]{C}{1}{1C}{-4.5}{0}
\linkedeffect[0.7]{C}{2}{2C}{-3}{0}
\linkedprep[0.7]{C}{1}{C1}{3}{0}
\linkedprep[0.7]{C}{2}{C2}{4.5}{0}
\end{move}
\end{diagram}
}
\begin{verbatim}
\begin{diagram}
\Duobox[2]{C}{0,0}
\linkedeffect[0.7]{C}{1}{1C}{-4.5}{0}
\linkedeffect[0.7]{C}{2}{2C}{-3}{0}
\linkedprep[0.7]{C}{1}{C1}{3}{0}
\linkedprep[0.7]{C}{2}{C2}{4.5}{0}
\end{diagram}
\end{verbatim}

\noindent We can connect wires to these fiducial elements using the names they are given (e.g.\ \verb+1C+ for the first one).

The following objects can be linked to (like duoboxes) and have symbols placed above (or below) them
\makebox[3cm][l]{
\begin{diagram}
\boundingbox{0,0}{0,0}
\begin{move}{23,-3.5}
\bbmetric{A}{0,0}
\bwmetric{B}{0,-2}
\wbmetric{C}{0,-4}
\wwmetric{D}{0,-6}\otherside \duosymbol{d}
\end{move}
\end{diagram}
}
\begin{verbatim}
\begin{diagram}
\bbmetric{A}{0,0}
\bwmetric{B}{0,-2}
\wbmetric{C}{0,-4}
\wwmetric{D}{0,-6}\otherside \duosymbol{d}
\end{diagram}
\end{verbatim}

\noindent We have also shown how to put a symbol below one of these objects.

To foliate a circuit we use the \verb+foliation+ environment.  This contains an argument saying what $x$-coordinate on the left the foliation lines start, and what $x$-coordinate on the right the foliation lines end.  Foliation lines are drawn using \verb+\startfoliate+, \verb+\continuefoliate+, and \verb+\finishfoliate+ commands.  The \verb+\Startfoliate+, \verb+\Continuefoliate+, and \verb+\Finishfoliate+ commands take an extra argument which allow the intersection point of foliation line with a given wire to be moved away from the centre of the wire by a fraction of the wires length.
\makebox[3cm][l]{
\begin{diagram}[1.5]
\boundingbox{0,0}{0,0}
\begin{move}{18,-15}
\Opbox{A}{0,0}
\Opbox[2]{B}{2,4}
\Opbox[2]{C}{-1,9}
\wire{A}{B}{2}{1}
\wire{A}{B}{3}{2}
\wire{A}{C}{1}{1}
\wire{B}{C}{1.5}{2}
\begin{foliation}{-5}{6}
\Startfoliate{A}{C}{1}{1}{-0.25}
\continuefoliate{A}{B}{2}{1}
\finishfoliate{A}{B}{3}{2}
\putlatex{\ensuremath{t_1}}
\Startfoliate{A}{C}{1}{1}{0.25}
\finishfoliate{B}{C}{1.5}{2}
\otherside\putlatex{\ensuremath{t_1}}
\end{foliation}
\end{move}
\end{diagram}
}
\begin{verbatim}
\begin{diagram}[1.5]
\Opbox{A}{0,0}
\Opbox[2]{B}{2,4}
\Opbox[2]{C}{-1,9}
\wire{A}{B}{2}{1}
\wire{A}{B}{3}{2}
\wire{A}{C}{1}{1}
\wire{B}{C}{1.5}{2}
\begin{foliation}{-5}{6}
  \Startfoliate{A}{C}{1}{1}{-0.25}
  \continuefoliate{A}{B}{2}{1}
  \finishfoliate{A}{B}{3}{2}
  \putlatex{\ensuremath{t_1}}
  \Startfoliate{A}{C}{1}{1}{0.25}
  \finishfoliate{B}{C}{1.5}{2}
  \otherside\putlatex{\ensuremath{t_1}}
\end{foliation}
\end{diagram}
\end{verbatim}

\noindent We have illustrated a few more features in this example.  The \verb+diagram+ environment takes an optional argument that scales the diagram (in this example we have set this to 1.5, the default is 1).  Scaling does not effect the font size.  We have also used the \verb+\putlatex+ command.  This works like \verb+\opsymbol+ and \verb+\duosymbol+. It places the symbol inside the argument at the appropriate location (in this case on the left of the foliation line, or on the right if the \verb+\otherside+ command is used).  We have used the standard LaTeX command \verb+\ensuremath+ rather than \verb+$...$+  since then the diagram can be put inside the equation environment without causing an error.  The command \verb+\placelatex{x,y}{LaTeX}+ can also be used to put some standard LaTex in the diagram.  In this case, it places it at the position \verb+(x,y)+.  Both \verb+\putlatex+ and \verb+\placelatex+ admit an optional fine tuning argument.

The command \verb+\thispoint[dx,dy]{A}{x,y}+ produces an abstract point named \verb+A+ which we can connect wires to, links to, and place symbols at using the various commands above.  This can be useful if we want a wire or link to end at a location at which there is no box.

We can use \verb+\begin{Diagram}[scale]{X}{y} ... \end{Diagram}+ instead of using the \verb+diagram+ environment.  This has additional arguments.  \verb+X+ moves the whole diagram to the left by this distance.  \verb+y+ changes the baseline LaTeX associated with the diagram which has the effect of moving the diagram up by this distance in some cases (this is useful in the equation environment).   Within a diagram we can group together different parts of the diagram using the \verb+move+ environment.  Thus, the part of the diagram whose commands appear inside \verb+\begin{move}[scale]{X,Y} ... \end{move}+ will be scaled by a factor \verb+scale+ and moved distance \verb+X+ in the $x$-direction and \verb+Y+ in the $y$-direction relative to other parts of the diagram.   This is useful in managing complex diagrams.
The \verb+\boundingbox{x1,y1}{x2,y2}+ command causes LaTeX to think that the diagram is inside the bounding box with bottom left corner at \verb+(x1,y1)+ and top right corner \verb+(x2,y2)+.  This is useful in managing the white space around a diagram.

\verb+duotenzor+ provides a number of stand alone commands.  These can be used in regular text, in math mode, and in the equation environment but not in the \verb+diagram+ environment.  These commands are

\newpage

\hspace{2.2cm}\makebox[\textwidth][l]{Stand alone commands in duotenzor}

\begin{center}
\begin{tabular}{l l c l l}
\bbdots & \verb+\bbdots+ & \hspace{3cm} & \bndots & \verb+\bndots+  \\
\bwdots & \verb+\bwdots+ & & \wndots & \verb+\wndots+  \\
\wbdots & \verb+\wbdots+ & & \nbdots & \verb+\nbdots+  \\
\wwdots & \verb+\wwdots+ & & \nwdots & \verb+\nwdots+  \\
\nndots & \verb+\nndots+ & \vspace{5pt} & \nndotslong & \verb+\nndotslong+ \\
\bbleftjump & \verb+\bbleftjump+ & \vspace{6pt} & \bbrightjump & \verb+\bbrightjump+  \\
\bwleftjump & \verb+\bwleftjump+ & \vspace{6pt} & \bwrightjump & \verb+\bwrightjump+ \\
\wbleftjump & \verb+\wbleftjump+ & \vspace{6pt} & \wbrightjump & \verb+\wbrightjump+ \\
\wwleftjump & \verb+\wwleftjump+ & \vspace{6pt} & \wwrightjump & \verb+\wwrightjump+ \\
\end{tabular}
\end{center}

One more stand alone command is \verb+\ultrathickdash+ which produces \ultrathickdash.
Finally, for reference, here is a complete list of all commands (in addition to those in the above table) most of which have been discussed above.  They are grouped with other commands (after \lq\lq also") which take similar arguments.  The argument in square brackets is always optional.  For \verb+step+, \verb+scale+, \verb+size+, \verb+curviness+, and \verb+bendiness+  the default value is 1.  For the fine tuning parameters \verb+dx,dy+ the default value is $0,0$. The \verb+\setoperatinalfont+ command can be used to change the operational font. It is set, by default to \verb+\mathsf+. The \verb+\constructiongrid+ command can be used to aid drawing (and removed from the finished product). Its arguments indicate the lower left and upper right coordinates.

\begin{verbatim}
Fonts:     \setoperationalfont{\newfontchoice},
                also \setduotensorfont{\newfontchoice},
Environ-:  \begin{diagram}[scale],
ments      \begin{Diagram}[scale]{xshift}{voffset},
           \begin{move}[scale]{X,Y},
Grid:      \constructiongrid[step]{x1,y1}{x2,y2}
Boxes:     \Opbox[size]{A}{x,y},
                also \opbox, \Duobox, \duobox,
           \Opduobox{hsize}{vsize}{A}{x,y}
                also \opduobox
Fiducials: \fidprep[size]{F}{x,y},
                also \fideffect
           \linkedprep[size]{A}{3}{X}{x}{Dy},
                also \linkedeffect
Wires:     \wire[curviness]{A}{B}{1}{3},
                also \doublewire, \thickwire, \thinwire
           \outwire[bendiness]{A}{2},
                also \inwire, \closedinwire, \closedoutwire
Links:     \link[curviness]{A}{B}{1}{3},
                also \doublelink, \thicklink, \thinlink, \linkbw,
                     \linkwb, \linkww, \linkbb, \blink, \linkb,
                     \wlink, \linkw, \linkwbbw, \linkbwwb,
                     \linkwbwb, \linkbwbw, \blink, \wlink, \linkb
                     \linkw, \blinkb, \blinkw, \wlinkb, \wlinkw
           \outblack[bendiness]{A}{3},
                also \outwhite, \inblack, \inwhite
Symbols:   \opsymbol[dx,dy]{a},
                also \duosymbol, \Opsymbol, \Duosymbol, \putlatex
           \otherside
           \placelatex[dx,dy]{x,y}{LaTeX stuff here}
Metric:    \bbmetric[size]{g}{x,y}
                also \wwmetric, \wbmetric, \bwmetric
                     \bbinsert, \wbinsert, \bwinsert, \wwinsert
Foliation: \begin{foliation}{X}{Y}  ...  \end{foliation}
           \startfoliate{A}{B}{1}{3}
                also \continuefoliate, \finishfoliate
           \Startfoliate{A}{B}{1}{3}{Dy}
                also \Continuefoliate, \Finishfoliate
Abstract:  \thispoint[dx,dy]{A}{x,y}
           \boundingbox{x1,y1}{x2,y2}
\end{verbatim}


\begin{thebibliography}{00}
\bibitem{Flocality} L.\ Hardy, Formalism Locality in Quantum Theory and Quantum Gravity, arXiv:0804.0054 (2008). 
\bibitem{Penrosenotation}R.\ Penrose, Applications of negative dimensional tensors. In D. J. A. Welsh, editor,
Combinatorial Mathematics and its Applications, pages 221-244. Academic
Press, New York, (1971).
\bibitem{roadtoreality} R.\ Penrose, The road to reality: a complete guide to the laws of the universe, Vintage Books, (2005).  
\bibitem{AbramskyCoecke} S. Abramsky and B. Coecke. A categorical semantics of quantum protocols.
\textit{Proceedings of the 19th Annual IEEE Symposium on Logic in Computer Science
(LICS '04)}, pages 415-425 (2004).
\bibitem{quantumpicturalism} B.\ Coecke, Quantum picturalism, Contemporary Physics, \textbf{51}, pages 59-83 (2010).
\bibitem{foliable} L.\ Hardy, Foliable operational structures for general probabilistic theories, arXiv:0912.4740 (2009).
\bibitem{Selinger} P. Selinger, A survey of graphical languages for monoidal categories. In \textit{New
Structures for Physics}, B. Coecke (ed), pages 275–337, Springer-Verlag (2009).
\bibitem{fiveaxioms} L.\ Hardy, Quantum theory from five reasonable axioms,  arXiv:quant-ph/0101012 (2001).
\bibitem{causaloid1}L. Hardy. Probability theories with dynamic causal structure: A new framework for quantum gravity,  arXiv:gr-qc/0509120 (2005). 
\bibitem{Mackey} G.\ Mackey, \textit{Mathematical Foundations of Quantum Mechanics} (Benjamin, 1963).
\bibitem{Ludwig} G.\ Ludwig, \textit{An Axiomatic Basis of Quantum Mechanics}, volumes 1 and 2 (Springer-Verlag, 1985, 1987).
\bibitem{DaviesLewis} E.\ B.\ Davies and J.\ T.\ Lewis, An operational approach
to quantum probability, Communications in Mathematical Physics, \textbf{17}, pages 239-260 (1970).
\bibitem{Wootters1} W.\ K.\ Wootters, Quantum mechanics without probability amplitudes,  Foundations of Physics, \textbf{16}, 391 (1986).
\bibitem{FoulisRandall} D.\ J. Foulis and C.\ H.\ Randall, Empirical logic and tensor products. In \textit{Interpretations and Foundations of Quantum Theory}, edited by H.~Neumann (Bibliographisches Institut, Wissenschaftsverlag, Mannheim, 1981).
\bibitem{Gudder} S.\ Gudder, S.\ Pulmannov\'a, S.\ Bugajski, and
E.~Beltrametti. Convex and linear effect algebras.
Reports on Mathematical Physics, \textbf{44}, pages 359-379 (1999).
\bibitem{Barrett} J.\ Barrett, Information processing in generalized probabilistic theories, Physical
Review A, \textbf{75}  032304 (2007).
\bibitem{CDP} G.\ Chiribella, G.\ M.\ D'Ariano, and P.\ Perinotti, Probabilistic theories with purification. arXiv:0908.1583 (2009).
\bibitem{BarnumWilce} H.\ Barnum and A.\ Wilce, Information processing in convex operational theories, arXiv:0908.2352 (2009).    
%
\bibitem{causaloid2}L.\ Hardy Towards quantum gravity: A framework for probabilistic theories with non-fixed causal structure. Journal of Physics, \textbf{A40}, pages  3081-3099 (2007).
\bibitem{causaloid3} L. Hardy, Quantum gravity computers: On the theory of computation with indefinite causal structure, in {\it Quantum Reality, Relativistic Causality, and Closing the Epistemic Circle - Essays in Honour of Abner Shimony}, The Western Ontario series in philosophy of science 73, (Springer Science+Media B.V.\ 2009)   
\bibitem{causaloid4} S.\ Markes and L.\ Hardy, Entropy for theories with indefinite causal structure, arXiv:0910.1323 (2009).  
\bibitem{TS} S Tomonaga, On a relativistically invariant formulation of the quantum theory of wave
fields, Progress of Theoretical Physics \textbf{1}, 27 (1946).
\bibitem{TS2} J Schwinger, Quantum electrodynamics I. A covariant
formulation, Physical Review 74, 1439 (1948).
\bibitem{Fotini} F. Markopoulou. Quantum causal histories. Classical and Quantum Gravity, \textbf{17}, pages 2059–2077 (2000).
\bibitem{BIP} R.F. Blute, I.T. Ivanov, and P. Panangaden. Disgrete quantum causal dynamics.  International Journal of Theoretical Physics \textbf{42}, pages 2025-2041 (2003).
\bibitem{DAriano} G.\ M.\ D'Ariano, On the "principle of the quantumness", the quantumness of Relativity, and the computational grand-unification, arXiv:1001.1088 (2010).
\bibitem{ABL} Y. Aharonov, P. G.Bergmann and J. Lebowitz, Time symmetry in the quantum process of measurement, Physical Review  
134, B1410 (1964).
\bibitem{Aharonovetal} Y. Aharonov, S. Popescu, J. Tollaksen, L. Vaidman, Multiple-time states and multiple-time measurements in quantum mechanics, arXiv:0712.0320 (2007).  
\bibitem{Oeckl} R.\ Oeckl, General boundary quantum field theory: Foundations and probability interpretation, Advances in Theoretical and Mathematical  Physics, 12, pages 319-352 (2008).
\bibitem{Segal} G. Segal, The definition of conformal field theory, Differential geometrical
methods in theoretical physics (Como, 1987), Kluwer, Dordrecht, pp. 165–171 (1988).
\bibitem{Atiyah} M.\ Atiyah, Topological quantum field theories, Publications Mathématiques de l'IHÉS 68 (68): 175–186, (1988). 
\bibitem{strategies} G.\ Gutoski and J.\ Watrous, Towards a general theory of quantum games,  arXiv:quant-ph/0611234 (2006)
\bibitem{combs} G.\ Chiribella, G.\ M.\ D'Ariano, P.\ Perinotti, Quantum Circuits Architecture, arXiv:0712.1325 (2007).
\bibitem{EricBob} B.\ Coecke, E.\ O.\ Paquette, Categories for the practising physicist, arXiv:0905.3010 (2009).
\bibitem{paviapictures} G.\ Chiribella, G.\ M.\ D'Ariano, P.\ Perinotti, Probabilistic theories with purification,  arXiv:0908.1583 (2009). 
 \bibitem{Araki} H.~Araki, On a Characterization of the State Space of Quantum Mechanics.  Communications
in Mathematical Physics \textbf{75}, 1 (1980).
\bibitem{Bergia} S.~Bergia, F.~Cannata, A.~Cornia, and R.~Livi, On the actual measurability of the density matrix of a decaying system by means of measurements on the decay products.  Foundations of Physics \textbf{10}, 723 (1980).
\bibitem{Wootters86} W. K. Wootters, Local accessibility of quantum states.  In \textit{Complexity, Entropy and the Physics
of Information}, edited by W. H. Zurek (Addison-Wesley, 1990).
\bibitem{Mermin} N.~D.~Mermin, What is quantum mechanics trying to tell us?  American Journal of Physics \textbf{66}, 753 (1998).
\bibitem{HardyWootters} L.\ Hardy and W.\ K.\ Wootters, Limited Holism and Real-Vector-Space Quantum Theory, arXiv:1005.4870 (2010).  
 \bibitem{TikZ} T. Tantau, TikZ and PGF: Manual for version 2.00, at
http://sourceforge.net/projects/pgf/ (2007).
\bibitem{CTAN}L.\ Hardy, The duotenzor drawing package, at 
 
\noindent http://tug.ctan.org/tex-archive/graphics/duotenzor/ ~(2010).
\end{thebibliography}
\end{document}